\documentclass[%
                amsmath,
                twocolumn]{aastex702}
\usepackage[normalem]{ulem}
\usepackage{siunitx}
\DeclareSIUnit{\Msun}{M_{\odot}}
\DeclareSIUnit\erg{erg}

\definecolor{orange}{rgb}{1.0, 0.5, 0.0}

\begin{document}

\title{Strangeness Transport in Binary Neutron Star Mergers}

\author[0009-0002-2008-1250]{Melvin Storbacka}
\email[show]{mstorbac@caltech.edu}
\affiliation{TAPIR, Mailcode 350-17, California Institute of Technology, Pasadena, CA 91125, USA}
\affiliation{Department of Physics, KTH Royal Institute of Technology, SE-106 91, Stockholm, Sweden}

\author[0000-0003-3829-967X]{Jiaxi Wu}
\email[]{jiaxiwu@caltech.edu}
\affiliation{TAPIR, Mailcode 350-17, California Institute of Technology, Pasadena, CA 91125, USA}

\author[0000-0002-5511-9565]{Alexander Haber}
\email{a.haber@soton.ac.uk}
\affiliation{Mathematical Sciences and STAG Research Centre,
University of Southampton, Southampton SO17 1BJ, United Kingdom}

\author[0000-0002-0491-1210]{Elias R. Most}
\email[]{emost@caltech.edu}
\affiliation{TAPIR, Mailcode 350-17, California Institute of Technology, Pasadena, CA 91125, USA}
\affiliation{Walter Burke Institute for Theoretical Physics, California Institute of Technology, Pasadena, CA 91125, USA}

\author[0000-0003-3229-4958]{Jacquelyn Noronha-Hostler}
\email[]{jnorhos@illinois.edu}
\affiliation{Illinois Center for Advanced Studies of the Universe, Department of Physics, University of Illinois at Urbana-Champaign, Urbana, IL 61801, USA}

\author[0000-0002-2189-706X]{Mateus Reinke Pelicer}
\email[]{matrp@illinois.edu}
\affiliation{Illinois Center for Advanced Studies of the Universe, Department of Physics, University of Illinois at Urbana-Champaign, Urbana, IL 61801, USA}

\author[0009-0004-7870-0039]{Nikolas Cruz-Camacho}
\email[]{cnc6@illinois.edu}
\affiliation{Illinois Center for Advanced Studies of the Universe, Department of Physics, University of Illinois at Urbana-Champaign, Urbana, IL 61801, USA}

\author[0000-0001-5578-2626]{Veronica Dexheimer}
\email[]{vdexheim@kent.edu}
\affiliation{Department of Physics, Kent State University, Kent, OH 44243, USA}
\begin{abstract}
The presence of hyperons in the cores of neutron stars opens fast
{strangeness equilibration }channels that can produce bulk-viscous dissipation during binary inspiral. Because these reactions coexist with {electron} $\beta$-equilibration, tidal compression can drive {the} two coupled chemical imbalances far beyond linear response. We construct the first reaction network that self-consistently evolves the electron and strangeness fractions with a four-dimensional {strangeness-dependent} chiral mean-field {(CMF)} equation of state, including nucleonic and hyperonic Urca processes and non-leptonic hyperon reactions. For periodic density perturbations, representative of inspiral oscillations, we find that rapid strangeness conversion can generate a large $\beta$-imbalance, after which slow $\beta$-equilibration bottlenecks strangeness relaxation. {Rather than decaying exponentially,} the coupled system consequently exhibits {dynamically important} algebraic {decay in a far-from-equilibrium regime}. At the $\rm keV$ temperatures expected during inspiral, this nonlinear response produces a broad enhancement of the effective bulk viscosity, reaching $\sim10^{31}\,\mathrm{g\,cm^{-1}\,s^{-1}}$ for $320$ Hz oscillations. A phenomenological estimate of continuous inspiral dissipation yields gravitational-wave phase shifts up to $\sim0.14$ rad for neutron stars with hyperonic cores. Self-consistent, far-from-equilibrium strangeness transport may therefore provide a dynamical probe of hyperons in neutron-star interiors.

\end{abstract}

\keywords{Neutron star cores (1107), Neutron stars (1108), Nuclear astrophysics (1129), Nuclear physics (2077), Gravitational waves (678) }

\section{Introduction}
Probing extreme states of matter {with neutron stars} is {currently} one of the fundamental goals of physics
and astrophysics.
Their extreme densities beyond nuclear saturation and high compactness make
them ideal probes for using macroscopic properties of stellar structure to
constrain the nuclear matter conditions inside them \citep{Lattimer:2000nx}.
{In particular, physicists are interested in the possibility of exotic particles and states of matter that may exist within their cores, such as strange particles like hyperons \citep{Gal:2016boi,Tolos:2020aln}. 
Nuclear physics experiments and chiral effective field theory provide constraints at densities around nuclear saturation (e.g. the inner crust) \citep{Drischler:2021kxf,MUSES:2023hyz,Tsang:2023vhh}, and perturbative quantum chromodynamics provides constraints at densities 1-2 orders of magnitude larger than the maximum central densities \citep{Komoltsev:2021jzg}, but the matter at the core of neutron stars remains a mystery and requires astrophysical constraints. }

{Several} avenues exist to constrain bulk properties of neutron stars, e.g., 
through pulsar timing \citep{Antoniadis:2013pzd,NANOGrav:2019jur,Fonseca:2021wxt}, X-ray emissions from hot spots due to
magnetospheric return currents \citep{Watts:2016uzu,Ozel:2016oaf,Miller:2019cac,Miller:2021qha,Riley:2019yda,Riley:2021pdl}, and gravitational wave observations
\citep{Flanagan:2007ix,Hinderer:2016eia,Chatziioannou:2021tdi, Hammond:2025kki},
{which when combined together can provide strong constraints on the underlying equation of state (EOS) (see, e.g., \citet{MUSES:2023hyz} for a recent review).} 

The first observation of the gravitational wave event GW170817 of
two merging neutron stars has already led to several constraints on the
dense matter EOS via imprints of the adiabatic tidal deformability on the last orbits of the inspiral (e.g., \citet{Dexheimer:2018dhb,Chatziioannou:2019yko,Chatziioannou:2018vzf,LIGOScientific:2018cki,Most:2018hfd,Annala:2017llu,Raithel:2018ncd}). Additional, more model
dependent constraints can be inferred by assuming the outcome of the merger
remnant \citep{Baiotti:2016qnr}, and using that to constrain the maximum mass of
neutron stars \citep{Margalit:2017dij,Rezzolla:2017aly,Ruiz:2017due,Shibata:2019ctb}. Additional constraints can be derived
using the second (and more massive) neutron star merger event
GW190425 \citep{LIGOScientific:2020aai} and its potential close connection to prompt black hole formation
\citep{Bauswein:2020aag,Kashyap:2021wzs,Tootle:2021umi}.

{While useful, these constraints cannot yet confirm or rule out specific internal compositions of the core \citep{Mroczek:2023zxo}.} This is because there are intrinsic
degeneracies between the EOS and the tidal deformability
\citep{Gamba:2019kwu,Tews:2018iwm,Legred:2023als,Essick:2023fso,Raithel:2022efm,Raithel:2022aee} (though there is strong quasi-universality, e.g.,  \citet{Yagi:2013awa,Yagi:2013bca,Yagi:2015pkc,Godzieba:2020bbz}), and between
different nuclear compositions entering the EOS and the
resulting observable bulk properties.
This means that even with next-generation facilities, fundamental questions
such as the presence of a deconfinement phase transition \citep{Alford:2013aca,Chatziioannou:2019yko} cannot
for all models be determined uniquely from observations of the tidal
deformability alone \citep{Raithel:2022efm,Essick:2023fso,Clevinger:2025acg}. 

Consequently, other potential non-adiabatic
imprints have been proposed, in particular associated with microphysical dissipation.
Matter inside the neutron star is subject to both strong and weak
interactions \citep{Shapiro:1983du,Yakovlev:2000jp}. Akin to a chemical equilibrium system, Urca processes
\citep{Gamow:1941gis,Lattimer:1991ib} regulate the relative lepton fraction so that matter tends
towards weak-interaction equilibrium. However, because of compression
present during the inspiral, and in particular during resonances, e.g., for
the g-mode \citep{1992ApJ...395..240R,1994ApJ...426..688R,laiResonantOscillationsTidal1994,Counsell:2024pua,Andersson:2025iyd,Zhao:2025pgx, Ghosh:2026ldg}, matter can be driven out of chemical equilibrium
\citep{1992ApJ...395..240R}. For matter containing strangeness, e.g., hyperons \citep{Schaffner:1995th},
chemical equilibration not only affects the lepton chemical potential, but
also the strangeness one. The impact of chemical equilibration is
that any multi-component system driven back to equilibrium will behave as an
effective viscous fluid \citep{gavassinoRelativisticBulkViscous2023,hernandezBurgersEquationBulk2025}. Viscosity here is to be interpreted in a
far-from-equilibrium sense, where the bulk scalar pressure can be large and
have a (transient) non-zero equilibrium value \citep{gavassinoRelativisticBulkRheology2024}. Depending on the
effective phase lag of the driving and the amplitude of the oscillations,
bulk viscosity from the system can become very large \citep{Yang:2023ogo}, and
potentially affect the inspiral dynamics \citep{Alford:2017rxf,Most:2021zvc,ripleyProbingInternalDissipative2023}. 
{For npe matter, both the bulk viscosity and its relaxation time can depend sensitively on the nuclear symmetry energy, with variations of its slope
changing the viscosity by orders of magnitude \citep{Yang:2025yoo}.}

While this phenomenon has
been analyzed using different approaches, including Navier-Stokes hydrodynamics
\citep{Celora:2022nbp} or relativistic first-order hydrodynamics \citep{ripleyProbingInternalDissipative2023,r.DissipativeTidalEffects2024}, the general
conclusion for leptonic dissipation is that the bulk viscosity may be too
small to leave a dynamical impact on the gravitational wave phase in the
inspiral itself \citep{Alford:2023gxq,HegadeKR:2026iou}. However, when strangeness equilibration is
included, in particular in the form of hyperonic matter, the bulk viscosity
can be orders of magnitude larger when temperatures are substantially
below $1\, \rm MeV$ \citep{Arras:2018fxj}. Therefore, it has been argued that exotic
phases of matter might be detectable with next-generation gravitational
wave facilities \citep{Yu:2017cxe,alfordStrangenesschangingRatesHyperonic2021,ghoshTidalHeatingDirect2024}, especially if oscillations of the neutron star are
resonantly excited, e.g., at the g-mode or f-mode resonance \citep{1994ApJ...426..688R,laiResonantOscillationsTidal1994,Ghosh:2026ldg}.

{Because hyperonic matter seems to be the most promising {channel for producing dynamically important} bulk viscosity in the inspiral, self-consistent calculations and the inclusion of all coupled reactions {are crucial}. For instance, a single decay in isolation is not the same as a coupled, fully non-linear reaction network that is allowed to be far-from-equilibrium, which can significantly change equilibration times (see e.g. \cite{Noronha-Hostler:2007fzh,Noronha-Hostler:2009wof}).}
Previous studies of weak-interaction dissipation in dense matter have largely calculated an effective bulk viscosity from perturbations about chemical equilibrium, without evolving the underlying composition on a dynamical binary-merger background. For hyperonic matter, these calculations generally linearize the non-leptonic rates 
\citep{gusakovBulkViscositySuperfluid2008,ofengeimBulkViscosityNeutron2019,alfordStrangenesschangingRatesHyperonic2021}. 
More recent work has incorporated this linearized hyperonic relaxation into stellar-mode and tidal perturbation calculations \citep{Ghosh:2026ldg}, but still assumes a static stellar background and does not evolve the coupled $\beta$ and strangeness compositions.
The closest reaction-network precedent (similar to what we will perform here) instead comes from quark matter, where coupled non-leptonic and semi-leptonic reactions were treated first in linear response \citep{sadBulkViscosityStrange2007,hernandezBurgersEquationBulk2025} and subsequently with nonlinear reaction rates \citep{shovkovyBulkViscosityNonlinear2011}. 
{However, these quark matter reaction network calculations derived the bulk-viscous response of homogeneous matter under prescribed perturbations or near-equilibrium evolution, {also not taking into account oscillation conditions extracted from} dynamical numerical-relativity simulations.}

{Despite this progress, hyperonic bulk viscosity has not yet been studied using a self-consistent, coupled reaction network in the far-from-equilibrium regime relevant to binary inspiral.}
As we will show here, the inspiral phase probes exactly this  far-from-equilibrium regime, which is not properly described by simple bulk viscosity assumptions. 
{Additionally, } since both leptonic and non-leptonic rates can simultaneously convert strangeness, a fully self-consistent dynamical equilibration process by means of a full, coupled chemical reaction network is needed.

Here, we address this by presenting the first fully coupled strangeness reaction network
for use in neutron star merger calculations, enabled by the new MUSES EOS framework that allows us to fully self-consistently model strangeness dependence \citep{Jahan:2026hvs,pelicerBuildingNeutronStars2025}.
We describe the overall aim and summary of our results in Sec. \ref{sec:basicpicture}. We then provide details on our methodology in Sec. \ref{sec:methods}, before providing a detailed discussion of our results in Sec. \ref{sec:results}.

\section{Basic picture}
\label{sec:basicpicture}

In neutron stars, $\beta$-equilibration occurs through modified Urca processes $n + N \longrightarrow p + N + e^- + \bar\nu_e,\, p+N + e^- \longrightarrow n + N + \nu_e $, where $N$ is a spectator neutron or proton, and the nucleonic direct Urca processes $n \longrightarrow p + e^- + \bar\nu_e,\, p+e^- \longrightarrow n + \nu_e $ \citep{Lattimer:1991ib, Yakovlev:2000jp}. At very low temperatures, nucleonic direct Urca reactions are only possible {at proton fractions} $Y_p \gtrsim 0.11$ due to (Fermi) {energy and} momentum conservation in $npe$ matter, resulting in a minimum density threshold at which direct Urca begins to equilibrate the matter \citep{Lattimer:1991ib}. The Urca reactions impose the $\beta$-equilibrium condition, 
\begin{equation}
    \label{eq:betaeq}
    \mu_\beta := \mu_n - \mu_p - \mu_e + \mu_{\nu_e} = 0
\end{equation}
on the equilibrium state, where $\mu_i$ is the chemical potential of particle $i$, {and in the inspiral, $\mu_{\nu_e}\approx 0$, (whereas in the post-merger phase, neutrinos are trapped and $\mu_{\nu_e}$ is finite \citep{Espino:2023dei,Alford:2026kwd})}. {In the {low-temperature} multicomponent matter considered below, {only states near the Fermi surfaces are assumed to contribute to the reactions (see Appendix \ref{app:reactionrates})} {and thermal contributions, which are very small compared to the modified Urca channels at the considered temperatures, are neglected.}%

Apart from determining {$\beta$}-equilibrium, weak reactions may produce {and equilibrate} strangeness through, for example, the hyperonic direct Urca reactions, %
\begin{equation}
    \label{reac:strangeUrca}
    \Lambda \longrightarrow p + e^- + \bar\nu_e, ~~~ p + e^- \longrightarrow \Lambda + \nu_e,
\end{equation}
and the non-leptonic reactions, %
\begin{equation}
    \label{reac:non-leptonic}
    \begin{aligned}
        n + n &\longleftrightarrow n + \Lambda, \\
        n + p &\longleftrightarrow \Lambda + p, \\
        n + \Lambda &\longleftrightarrow \Lambda + \Lambda.
    \end{aligned}
\end{equation}
This introduces the additional equilibrium constraint,
\begin{align}
\mu_S ={ \mu_n - \mu_\Lambda = } 0\,,
\end{align}
on the strangeness chemical potential. {While other hyperons, including $\Xi$ and $\Sigma$, can potentially also be probed, $\Lambda$ is the dominant {hyperonic} species in the EOS we use (see Section \ref{subsec:EOS})}.

At low temperatures, the timescale of strangeness equilibration is much smaller than for $\beta$ equilibration \citep{vandalenBulkViscosityNeutron2004}. This is because the non-leptonic strangeness-changing reactions do not involve neutrinos and exhibit a $T^{3}$ dependence, while the direct (modified) Urca rates scale with $T^{5}$ ($T^{7}$), in part due to the temperature dependence of the phase-space of emitted neutrinos \citep{vandalenBulkViscosityNeutron2004}.

In equilibrium, the strangeness fraction, $Y_S = n_S/n_B$, and the electron fraction, $Y_e = n_e/n_B$, are determined by the baryon density $n_B$ and the energy density $\varepsilon$, with $n_S$ and $n_e$ denoting the strangeness and electron number densities, respectively. However, because of the finite timescale of equilibration, matter can be pushed out of chemical equilibrium. This gives rise to an out-of-equilibrium pressure correction, known as the bulk scalar $\Pi$, given by
\begin{equation}
\label{eq:bulkscalar}
    P(\varepsilon, n_B, Y_e, Y_S) = P_{\mathrm{eq}}(\varepsilon, n_B) + \Pi(\varepsilon, n_B, Y_e, Y_S),
\end{equation}
where $P$ is the pressure, and $P_{\mathrm{eq}}(\varepsilon, n_B) = P(\varepsilon, n_B, Y_e^\mathrm{eq}, Y_S^\mathrm{eq})$ is the pressure in chemical equilibrium. {Such a pressure correction can affect the dynamics of the fluid and lead to dissipation; thus, modeling the evolution of $\Pi$ is important for the description of the fluid.}

For matter with a single independent degree of freedom relaxing to equilibrium, it can be shown that the bulk scalar evolves according to {a resummed} Israel-Stewart-like equation \citep{gavassinoRelativisticBulkRheology2024, Yang:2023ogo}.
For systems with two independent relaxing degrees of freedom, the bulk scalar {does in general not have a closed form evolution equation \citep{gavassinoRelativisticBulkViscous2023}, and the system is instead governed through a multi-component relaxation of the chemical potentials $\mathbb{A}_a =(\mu_\beta, \mu_S)$,}
\begin{align}\label{eqn:mu_evol}
    &{  \mathcal{B}_{ab}(\mathbb{A})\; u^\mu\nabla_\mu \mathbb{A}^b}
   {+ \left(\delta^b_a - A^b S_a\right)\,\Gamma_b(\mathbb{A}) = -\,\mathcal{C}_a(\mathbb{A})\;\theta}\,,
\end{align}
{where $\Gamma_a$ is the reaction rate (see also Section \ref{sec:methods}), $u^\mu$ is the fluid four-velocity, ${\theta = }\nabla_\mu u^\mu$ the fluid expansion, $\mathcal{B}_{ab}(\mathbb{A})
   = -\,n_B \left.\frac{\partial Y_a}{\partial \mathbb{A}^b}
     \right|_{\mathfrak{s},n_B}$ the chemical susceptibility, $\mathcal{C}_a
   = n^2_B \left.\frac{\partial Y_a^{\rm eq}}{\partial n_B}\right|_{\mathfrak{s}}$ the compression coupling, $\mathfrak{s}$ is the entropy per baryon, $S^a= \frac{1}{T}\left.\frac{\partial Y_a}{\partial \mathfrak{s}}\right|_{n_B,\mathbb{A}}$ the reaction-heating susceptibility, and $a,b$ are chemical species.} {A full derivation is provided in Appendix \ref{app:burgers}}.
{In the limit where the rates become linear {in $\mathbb{A}_a$}, i.e., close to equilibrium, one can write $\Pi \simeq \mathcal{C}_a \mathbb{A}^a$, and Equation \eqref{eqn:mu_evol} reduces to Burgers-type equation }\citep{gavassinoRelativisticBulkViscous2023, hernandezBurgersEquationBulk2025},
\begin{align}\label{eqn:burgers}
    {\lambda u^\nu \nabla_\nu \left(u^\mu \nabla_\mu \Pi\right)
    + \tau_\Pi u^\mu \nabla_\mu \Pi  + \Pi = -\zeta \theta {- \chi} u^\nu\nabla_\nu \theta\,,}
\end{align}
{where $\lambda, \tau_\Pi, \chi$ are functions of the rates and EOS derivatives in Equation \eqref{eqn:mu_evol}, and $\zeta$ is called the bulk viscosity.}
In the limit of low-frequency oscillations, both the Israel-Stewart form and Burgers type behavior of {$\Pi$ reduce to the limit of a frequency-dependent AC bulk viscosity \citep{Yang:2023ogo}},
\begin{equation}
    \label{eq:NSlimit}
    \Pi \simeq -\zeta(\omega) \nabla_\mu u^\mu.
\end{equation}

In practical terms, solving Equation \eqref{eqn:mu_evol} directly is infeasible, and as we will show in general in the inspiral these reactions will fundamentally be in a (nonlinear) far-from-equilibrium regime \citep{gavassinoRelativisticBulkRheology2024}, where simple bulk viscous approaches, such as Eqs. \eqref{eqn:burgers} and \eqref{eq:NSlimit}, are inapplicable \citep{camelioSimulatingBulkViscosity2023, camelioSimulatingBulkViscosity2023a,Chabanov:2023abq}. As a result, the only way of accessing this regime is by directly solving the underlying rate equations directly, as we do in this work. In the case of treating $\beta-$equilibration only, this approach has been previously applied to simulations of post-merger neutron star dynamics \citep{Most:2022yhe}. 
Hyperonic matter instead
requires the simultaneous evolution of $Y_e$ and $Y_S$, with $\beta$- and
strangeness equilibration, which we formulate here.

\subsection{Impact on the inspiral}
During the inspiral of a binary neutron star merger, {each star} induces a tidal bulge in the {companion}. This tidal deformation drives fluid motion in the star, which can be damped by dissipative processes such as bulk or shear viscosity. This {effect} can siphon orbital energy into the fluid motion and subsequent heating of the star, resulting in a phase shift of the orbit compared to the non-dissipative case \citep{Bildsten:1992my,ripleyProbingInternalDissipative2023,Ghosh:2025glz}. 
\begin{figure}
    \centering
    \includegraphics[width=1.0\linewidth]{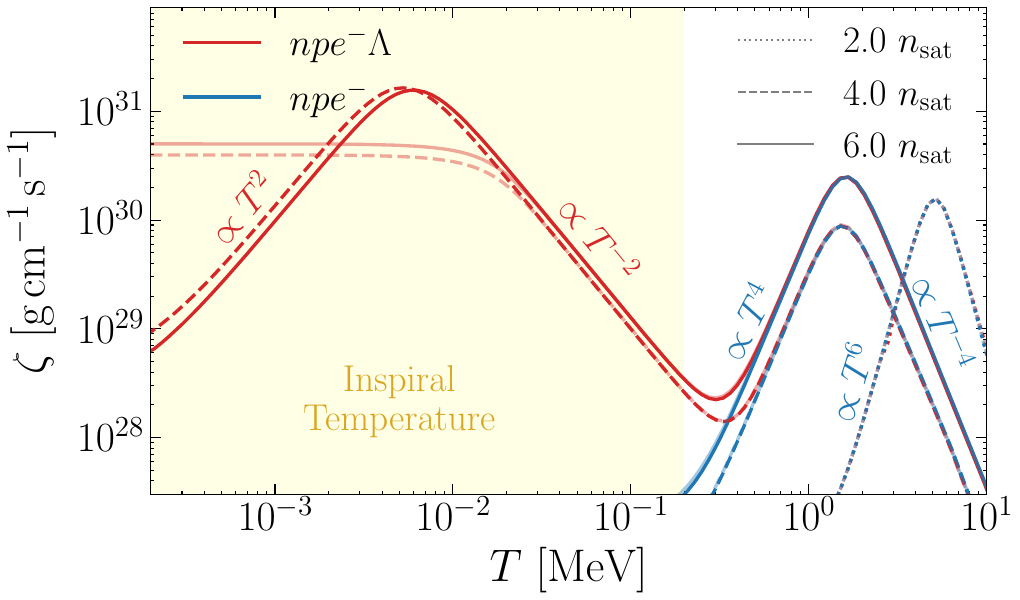}
    \caption{Bulk viscosity, $\zeta$, as a function of temperature, $T$, computed using our strangeness reaction network by tracking reactions in the fluid for $\SI{320}{\hertz}$ density oscillations around equilibrium. We include calculations with strangeness in ($npe^-$, blue) and out-of-equilibrium ($npe^-\Lambda$, red). 
    {{Full} lines show the results of very small density oscillations, while {transparent} lines show larger-amplitude oscillations of $\SI{0.5}{\percent}$.} The yellow-shaded area indicates temperatures expected during inspiral  \citep{Arras:2018fxj, ghoshTidalHeatingDirect2024}.}
    \label{fig:linearresponsebulk}
\end{figure}

\begin{figure*}
    \centering
    \includegraphics[width=0.8\linewidth]{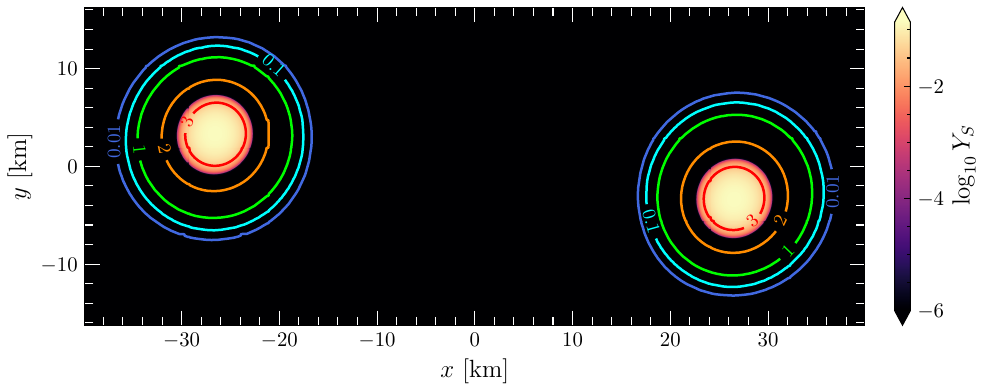}
    \caption{Spatial distribution of the strangeness fraction, $Y_S$, in an equal-mass inspiral of two $\SI{1.95}{\Msun}$ neutron stars, with the temperature fixed at $\SI{5}{\kilo\eV}$ (see Section \ref{subsec:inspiralsimulation}). Solid lines denote level contours of the baryon density in units of $n_\mathrm{sat}$.}
    \label{fig:inspiral}
\end{figure*}

In particular, the tidal forcing can resonantly excite neutron star g-mode oscillations that enhance the dissipative response \citep{HegadeKR:2026iou} and impart a sudden phase shift, which could be detectable down to $\Delta \Psi \simeq \SI{0.03}{\radian}$ for next-generation detectors \citep{Counsell:2024pua, gittinsDetectingTidalResonances2026}. Additionally, the resonances could yield sufficient fluid compression to heat the stars to $\sim\SI{0.01}{\mega\eV}$ from nucleonic Urca reactions alone \citep{laiResonantOscillationsTidal1994, Arras:2018fxj}.

At the low temperatures expected during inspiral, nucleonic bulk viscosity is expected to be subdominant to hyperonic bulk viscosity if hyperons are present (see Figure \ref{fig:linearresponsebulk}). Therefore, it has been suggested that the presence of a large phase shift from tidal dissipation could act as a probe of hyperons in the neutron star EOS \citep{Yu:2017cxe,ghoshTidalHeatingDirect2024, Ghosh:2025glz}.

In the case of the g-mode resonance, it could be significantly affected by fast chemical equilibration. At temperatures of a few MeV, the g-mode amplitude could be significantly decreased or completely suppressed by nucleonic Urca reactions \citep{ Zhao:2025pgx,HegadeKR:2026iou}.
During the inspiral, it is instead possible that hyperonic bulk viscosity suppresses these oscillations due to its faster equilibration rates \citep{alfordStrangenesschangingRatesHyperonic2021,Ghosh:2026ldg}.
To estimate the effect of hyperonic bulk viscosity on the g-mode\footnote{{Here, we estimate the effect of hyperonic bulk viscosity on g-modes caused by charge ($Y_e$) stratification, whereas \citet{Ghosh:2026ldg} consider g-modes caused by strangeness ($Y_S$) stratification.}}, the bulk viscous damping time of g-mode oscillations can be calculated as
\begin{equation}
    \frac{1}{\tau_g} = \frac{1}{2E_{g}}\int \zeta (n_B, T) \langle \theta ^2 \rangle dV.
\end{equation}
Assuming a g-mode frequency of $f_g = {\omega_g}/{2\pi} = \SI{320}{\hertz}$ and the region of strangeness to be concentrated to a core of radius $r_S \simeq\SI{6}{\kilo\meter}$ (see Figure \ref{fig:inspiral}) with approximately constant volume-averaged Lagrangian compression $k = \sqrt{\langle(\Delta \rho/\rho)^2\rangle}\simeq\SI{1}{\percent}$ and constant bulk viscosity $\zeta \simeq \SI{e31}{\gram\per\centi\meter\per\second}$ (see Figure \ref{fig:linearresponsebulk}), we find the approximate damping timescale for a $\SI{1.95}{\Msun}$ neutron star to be
\begin{equation}
\begin{aligned}
    \tau_g &\approx 2.3\cdot 10^{-3} \times \left(\frac{E_{g}}{\SI{2.1e48}{\erg}}\right)\left(\frac{\SI{6}{\kilo\meter}}{r_S}\right)^3 \\
    &\times \left(\frac{\SI{320}{\hertz}}{f_g}\right)^2 \left(\frac{0.01}{k}\right)^2\left(\frac{\SI{e31}{\gram\per\centi\meter\per\second}}{\zeta}\right)\,\si{\second},
\end{aligned}
\end{equation}
where the g-mode energy and estimates of the compression are taken from \citet{Arras:2018fxj}. Thus, $\omega_g\tau_g \simeq 4.6$, which indicates that the oscillation is underdamped, in which case we do not necessarily expect the g-modes to be suppressed by hyperonic bulk viscosity.\footnote{By the same argument, the nucleonic g-mode damping time can be estimated. For a warm neutron star with $T\simeq\SI{2}{\mega\eV}$, taking $\zeta_{npe}\simeq\SI{e30}{\gram\per\centi\meter\per\second}$ and estimating the compression to be $k\simeq4\%$ averaged over a core of $\sim\SI{9}{\kilo\meter}$ (the direct Urca core extent for a $\SI{1.95}{\Msun}$ neutron star with the EOS we use), we estimate $\tau_g\simeq\SI{0.43}{\milli\second}$. In this case, $\omega_g\tau_g \simeq 0.86$, yielding an overdamped mode, in alignment with expectations from \citet{Zhao:2025pgx}.} Instead, it is possible that the damping increases the deposited energy from the orbit, thereby enhancing the total phase shift. 

The above estimate depends sensitively on the g-mode core compression, energy, and frequency, {as well as the frequency regime over which g-mode oscillations can be driven \citep{Kwon:2024zyg,Kwon:2025zbc}}. Nonetheless, it provides a strong indication that hyperonic bulk viscosity is of importance to the energy transport during inspiral. In particular, we find that far-from-equilibrium strangeness transport will likely be required to properly model the hyperonic bulk viscosity across inspiral oscillations.

Having examined the sharp g-mode resonances, we now turn to the entire inspiral. During the inspiral, the dissipative part of the tidal response can be modeled using the (frequency-dependent) tidal dissipative deformability, $\Xi$. It enters the gravitational-wave phase at fourth post-Newtonian order away from resonances \citep{ripleyProbingInternalDissipative2023}, and provides a correction to the gravitational phase given by \citep{ripleyProbingInternalDissipative2023},
\begin{equation}
    \frac{d^2\Delta \Psi}{df^2} = 
    -\frac{225\pi^2}{3072}\left(\frac{GM}{c^3}\right)^2\Xi u^{-3}\,,
        \label{eq:GWphase}
\end{equation}
for a symmetric merger, where $u = (G\pi Mf/c^3)^{1/3}$, $M$ is the total binary mass, and $f$ is the GW frequency. Relating $\Xi$ phenomenologically to the volume-averaged bulk viscosity, $\langle \zeta \rangle$, through \citep{ripleyConstraintDissipativeTidal2024},
\begin{equation}
    \Xi = \frac{c^3}{G}\frac{p_{2}\Lambda}{C}\frac{\langle \zeta \rangle}{\langle \varepsilon \rangle M},
    \label{eq:dissipativedeformability}
\end{equation}
analysis of GW170817 has indicated upper bounds of $\Xi\lesssim 1200$ and $\langle \zeta \rangle \lesssim \SI{5e31}{\gram\per\centi\meter\per\second}$ \citep{ripleyConstraintDissipativeTidal2024, r.DissipativeTidalEffects2024}. Here, $C = \frac{GM_\mathrm{NS}}{Rc^2}$ is the compactness, and $\langle \varepsilon\rangle$ is the volume-averaged energy density. %
The factor $p_2$ is the tidal dissipative Love number that parametrizes the coupling between the tide and the bulk viscous dissipation, and remains unknown for realistic EOSs. However, from polytropic EOSs it has been found that $p_2$ ranges between $\sim0.002$ and $\sim 0.02$ \citep{HegadeKR:2024agt}. {For the $\SI{1.95}{\Msun}$ neutron star we consider here {and using our results on the bulk viscosity (see Figure \ref{fig:linearresponsebulk})}, we find,
\begin{equation}
    \Xi_{1.95} \simeq 8.1 \left(\frac{p_2}{0.01}\right)\left(\frac{\langle \zeta \rangle}{\SI{e31}{\gram\per\centi\meter\per\second}}\right),
\end{equation}
{The size of this number is in} large part due to the large compactness ($C \simeq 0.22$) and small tidal deformability ($\Lambda \simeq 64$) for this massive neutron star.}

 To estimate the phase shift caused by this continuous dissipation, we compute the frequency-dependent bulk viscosities of nucleonic equilibration and strange equilibration across the equilibrium structure of a neutron star {(see also Section \ref{sec:results})}. Using constant inspiral temperatures between $\SI{1}{\kilo\eV}$ and $\SI{10}{\kilo\eV}$, we propagate the bulk viscosities to a phase shift using Equation \eqref{eq:GWphase}, and find a qualitative difference between the phase shifts when hyperonic bulk viscosity is accounted for. The resulting phase shift from $f = \SI{20}{\hertz}$ until merger as a function of neutron star mass is shown in Figure \ref{fig:phase}, and the maximum phase shift we find is up to $\Delta \Psi \simeq \SI{0.14}{\radian}$. {While this} is sufficient to bias parameter estimation from the inspiral \citep{Ghosh:2025glz} and warrants incorporation into waveform modeling \citep{readWaveformUncertaintyQuantification2023}, {we caution that this likely presents an upper bound as the frequency-regime over which a large bulk viscosity can be maintained will depend strongly on the localization (or lack thereof \citep{Kwon:2024zyg}) of resonant modes \citep{HegadeKR:2026iou} and the resulting tidal response of the star \citep{HegadeKR:2024agt,Andersson:2025iyd}. }

\begin{figure}
    \centering
    \includegraphics[width=1.0\linewidth]{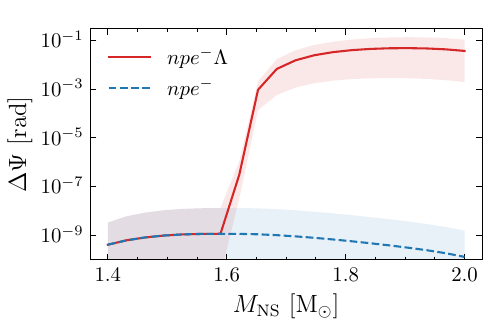}
    \caption{Gravitational wave phase shift between $f = \SI{20}{\hertz}$ and merger for a symmetric binary with component masses $M_\mathrm{NS}$, with and without hyperonic bulk viscosity. The lines correspond to mean phase shifts, $\Delta \Psi$, and the bands show the total range found by varying the temperature and tidal dissipative Love number, $p_2$. {Neutron star masses $M_{\rm NS}>1.6 M_\odot$ probe the onset of hyperons in the core.}}
    \label{fig:phase}
\end{figure}
\section{Methods}\label{sec:methods}
We probe the strangeness and $\beta$-equilibration dynamics of dense matter by tracking the coupled evolution of the particle fractions $Y_i$. Given a density evolution $n_B(t)$, the particle fractions evolve according to
\begin{equation}
    \label{eq:fractionevolution}
    u^\mu\nabla_\mu Y_i = \frac{1}{n_B}\Gamma_i,
\end{equation}
where $u^\mu$ is the fluid 4-velocity and $\Gamma_i$ is the total net volumetric rate of microphysical reactions affecting the particle fraction $Y_i$. The reaction rates are calculated by assuming that neutrinos are free-streaming 
and using the Fermi surface approximation (see Appendix \ref{app:reactionrates}). 

Over the timescales we consider, entropy changes, e.g., due to cooling, will be small \citep{Most:2022yhe}. As a result, we assume that heating as a result of the reactions can be neglected in Equation \eqref{eqn:app_entropy}, which implies that entropy per baryon, $\mathfrak{s}$ is well conserved along fluid streamlines, i.e., $u^\mu\nabla_\mu \mathfrak{s} = 0$, so that we can hold $\mathfrak{s}$ constant for each tracer, and then use this to recover the temperature dynamically throughout the evolution.
{In other words, we determine $\mathfrak{s}(T_0,n_B)$ from our initial temperature $T_0$ and $n_B$, but $T(t)$ is allowed to vary. However, over the density and composition range sampled by our tracers,  $\mathfrak{s}$ is expected to only vary weakly with $n_B$ at fixed temperature \citep{Mroczek:2024sfp}. Consequently, the relevant isotherms and isentropes nearly coincide, so our fixed-$\mathfrak{s}$ evolution is expected to produce only modest temperature variations such that $T_0\sim T(t)$.}
From the evolution of Equation \eqref{eq:fractionevolution}, the bulk scalar can be computed using Equation \eqref{eq:bulkscalar}. While this does not allow for self-consistent feedback of the out-of-equilibrium pressure on the fluid dynamics, it serves to probe the simultaneous equilibration of strangeness and $\beta$-processes, which constitutes the dominant effect. {An implementation of the reaction network into the simulation would allow us to consistently capture the effects of the coupled chemical equilibration and the resulting viscous effects. 
}

\begin{figure}
    \centering
    \includegraphics[width=1.0\linewidth]{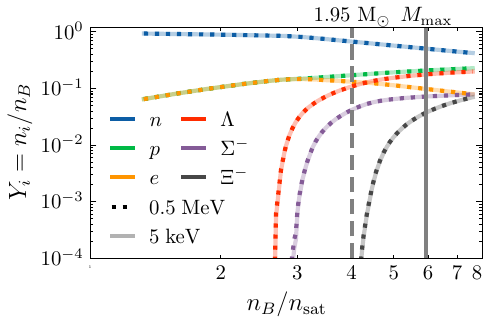}
    \caption{{Particle fractions, $Y_i$, probed for at different baryon number densities, $n_B$, for the C4 parameterization of the \texttt{CMF} EOS used in this work. Here, $n_{\rm sat}$ denotes nuclear saturation density. Gray vertical lines denote the central density of the $1.95\, M_\odot$ (dashed) and the maximum mass ($M_\mathrm{max} = 2.03\, M_\odot$) neutron star (solid) consistent with our EOS. Colored dotted and solid lines correspond to different temperature slices.}}
    \label{fig:EOSpopulation}
\end{figure}

Due to the large {numerical} %
stiffness arising from significant differences in equilibration times, we discretize Equation \eqref{eq:fractionevolution} using the implicit backward Euler method. With this discretization, the vector of particle number densities at time $j$, $\boldsymbol{n}_j$, is given by solving the equation
\begin{equation}
    \label{eq:discretized}
    \boldsymbol{n}_{j} = \Delta t \cdot \boldsymbol{\Gamma}_j(\varepsilon_j, n_{B, j}, \boldsymbol{n}_j) + \frac{n_{B,j}}{n_{B, j-1}}\boldsymbol{n}_{j-1}, 
\end{equation}
where $\Delta t$ is the timestep, $\boldsymbol{\Gamma}_j(\varepsilon_j, n_{B, j}, \boldsymbol{n}_j)$ is the vector of reaction rates computed at timestep $j$, and $\varepsilon_j$ and $n_{B,j}$ are the energy density and baryon number density at timestep $j$, respectively. %
The first term on the right hand side accounts for the effect of reactions, while the second term scales the number densities such that compression or expansion (without reactions) leaves the particle fractions $Y_i$ unchanged. 

The discretized Equation \eqref{eq:discretized} is evolved from the time evolution $n_B(t)$, an initial entropy per baryon, $\mathfrak{s}_0$, (or, equivalently, temperature or energy density), and the abundances from the initial number densities $\boldsymbol{n}_0 = \boldsymbol{n}^{\mathrm{eq}}(\mathfrak{s}_0, n_B(0))$. 
{A validation test and convergence analysis for the network are presented in Appendix \ref{subsec:validation}.}

\begin{figure*}
    \centering
    \includegraphics[width=.8\linewidth]{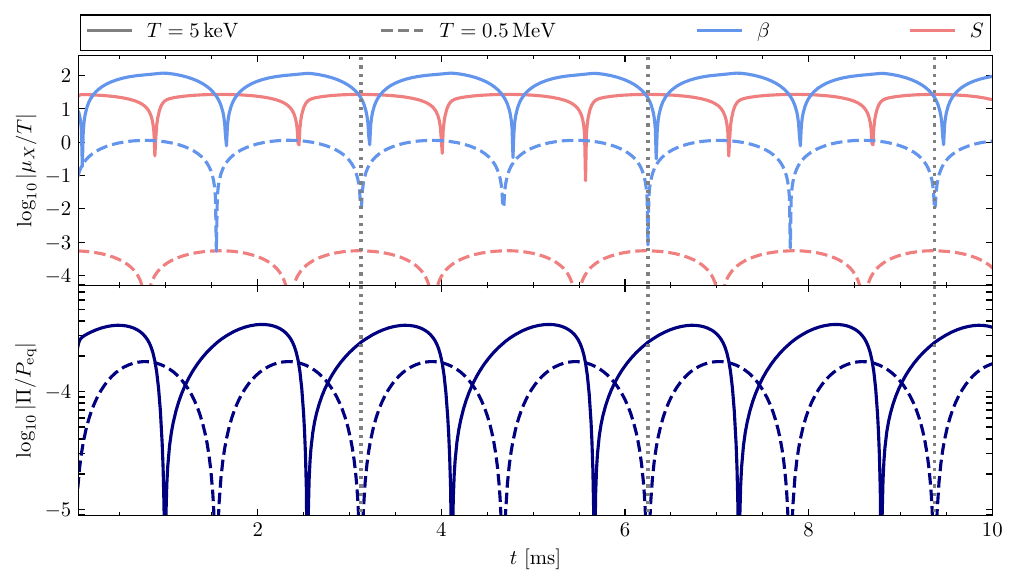}
    \caption{Evolution of chemical potential deviations, $\mu_X/T$, and the out-of-equilibrium pressure correction, $\Pi/P_\mathrm{eq}$, for $\SI{320}{\hertz}$ 1\% density oscillations around $3.5\,n_\mathrm{sat}$ computed using the strangeness reaction network. Dashed lines show the results for initial temperature $T=\SI{0.5}{\mega\eV}$, while solid lines show the case of $T=\SI{5}{\kilo\eV}$. {Here, $\beta$ and $S$ refer to the out-of-equilibrium chemical potentials for $\beta$- and strangeness processes, respectively.}
    }
    \label{fig:individualtracers}
\end{figure*}

\subsection{Equation of state}
\label{subsec:EOS}
To solve Equation \eqref{eq:fractionevolution}, an out-of-$\beta$-and-strangeness-equilibrium EOS depending on the energy density, baryon number density, and composition is required. We use the chiral mean-field (CMF) model computed with the \texttt{CMF++} code \cite{cruz-camachoPhaseStability3Dimensional2024}. The CMF model is an effective relativistic mean-field model based on a nonlinear realization of the $SU(3)$ sigma model with chiral symmetry restoration \citep{weinbergNonlinearRealizationsChiral1968, papazoglouNucleiChiralSU31999, Dexheimer:2008ax,cruz-camachoPhaseStability3Dimensional2024}. The model is consistent with both nuclear \citep{papazoglouNucleiChiralSU31999, Dexheimer:2008ax} and astrophysical constraints \citep{Dexheimer:2008ax, Dexheimer:2018dhb, dexheimerReconcilingNuclearAstrophysical2015} and is compatible with results from lattice \citep{dexheimerNovelApproachModeling2010} and perturbative \citep{roarkDeconfinementPhaseTransition2018} quantum chromodynamics. Details of the model and parameters can be found in \cite{cruz-camachoPhaseStability3Dimensional2024}. 

{In this work we make use of the C4 parametrization including the} full baryon octet ($n$, $p$, $\Lambda$, $\Sigma^-$, $\Sigma^0$, $\Sigma^+$, $\Xi^-$, $\Xi^0$) using the \textsc{CMF++} module within the MUSES cyberinfrastructure \citep{zenodo_cmf} 
and add electrons to fulfill local charge equilibrium using the Lepton module \citep{pelicer_2025_14654137}. 
{The particle fractions as a function of baryon density obtained using this EOS are shown in Figure \ref{fig:EOSpopulation}. %
The $\Lambda$ hyperon remains the dominant strangeness-carrying species over all densities we consider. Therefore, in Equation \eqref{eq:fractionevolution}, we evolve the fractions of neutrons, protons, electrons, and $\Lambda$ hyperons (see Appendix \ref{sec:rateequations}), and assume that the contributions of
the remaining hyperons to strangeness equilibration can be neglected. {This is based on their smaller} {contributions to the $\mu_S$ equilibration, which is largely a result of the smaller phase space of the contributing particles.\footnote{Compare Figures 1 and 3 of \citet{alfordStrangenesschangingRatesHyperonic2021}, in which the rate involving $\Sigma^-$ hyperons becomes subdominant when it is no longer the dominant hyperon.} Due to fast strong interaction channels \citep{ofengeimBulkViscosityNeutron2019, alfordStrangenesschangingRatesHyperonic2021}, three chemical potentials suffice to describe the densities of all baryons: $\mu_B$, $\mu_Q$, and $\mu_S$. {Thus, the evolution of the remaining hyperons} can be {determined} by calculating their density at the given chemical potentials and temperature.} {For the weak reaction rates entering Equation \eqref{eq:fractionevolution}, we}} include direct and modified Urca reactions, as well as the non-leptonic strangeness-changing reactions, with rates given in Appendices \ref{sec:dUrca}, \ref{sec:mUrca} and \ref{sec:non-leptonic}, respectively.

As the CMF EOS is only valid at densities above $n_\mathrm{sat}$ \citep{pelicerBuildingNeutronStars2025}, we match to the DD2 EOS \citep{typel_composition_2010, hempelStatisticalModelComplete2010} using the matching procedure of \citet{schneiderOpensourceNuclearEquation2017}.
With this combined EOS, hyperons first start appearing in the core for $T=0$ at a neutron star mass $\sim\SI{1.60}{\Msun}$, and the maximal neutron star mass is $\SI{2.03}{\Msun}$.

\subsection{Inspiral simulation}
\label{subsec:inspiralsimulation}
To probe which regimes are relevant for bulk viscosity and strangeness equilibration during the neutron star merger inspiral, we simulate a symmetric inspiral at an initial separation $\sim \SI{89}{\,\kilo\meter}$ using general-relativistic (magneto-)hydrodynamics \citep{Duez:2005sf} with spacetime evolution according to the Z4c formulation of the Einstein equations \citep{Hilditch:2012fp, Bernuzzi:2009ex}, as implemented in the \texttt{Frankfurt/IllinoisGRMHD (FIL)} code \citep{Most:2019kfe,Etienne:2015cea}. Initial data is computed using the \texttt{FUKA/Kadath} code \citep{Papenfort:2021hod,Grandclement:2009ju}. Details on the simulation setup can be found in \citet{Most:2022yhe}. We consider an equal-mass binary with component masses $M=\SI{1.95}{\Msun}$, in which the hyperonic regions extend $\sim\SI{6}{\kilo\meter}$ from the centers of the neutron stars. We extract density evolutions of co-moving fluid elements and evolve the composition of these using Equation \eqref{eq:fractionevolution}. We keep only fluid elements for which $n_B \gtrsim n_\mathrm{sat}$, as required for the network evolution using the CMF EOS. While this can provide an estimate of fluid oscillations in the inspiral, the temperature remains largely unconstrained due to numerical surface heating \citep{gittinsProblematicSystematicsNeutronstar2025}. Therefore, we cannot extract the initial temperature from our simulation, but instead prescribe an initial temperature $T=\SI{5}{\kilo\eV}$ from which the initial entropy is determined. 

\begin{figure*}
    \centering
    \includegraphics[width=.80\linewidth]{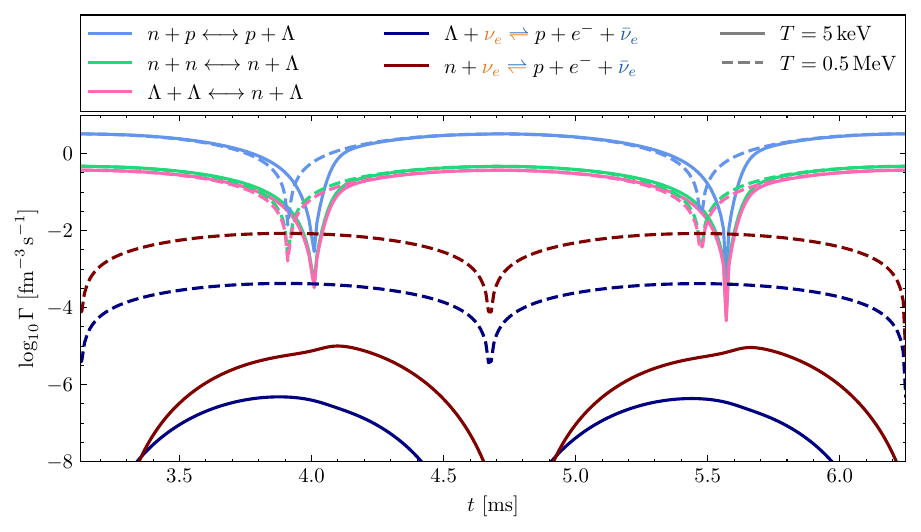}
    \caption{Evolution of individual reaction rates, $\Gamma$, over an oscillation cycle for the same oscillations as in Figure \ref{fig:individualtracers}. Modified Urca reactions are not shown because their reaction rates are at least $\sim 6$ orders of magnitude slower than the nucleonic direct Urca rates for the evolutions shown here. 
    }
    \label{fig:rates}
\end{figure*}

\section{Results}\label{sec:results}
We use the reaction network to probe equilibration and possible dissipative signatures in $npe\Lambda$ matter. We analyze the reaction rate dynamics of the equilibration and the regimes probed by the evolution in Section \ref{subsec:reactiondynamics}. In Section \ref{subsec:bulkviscosity}, we extract the bulk viscosity near and far-from equilibrium, and finally characterize the bulk viscosity and regimes probed during inspiral.

\subsection{Reaction dynamics}
\label{subsec:reactiondynamics}

Having validated the network and its convergence in Appendix \ref{subsec:validation}, we analyze the dynamics of the rates, chemical potential deviations, and the bulk scalar under forced sinusoidal oscillations. To do this, {we consider sinusoidal oscillations with parameter choices that reflect possible conditions during g-mode oscillations. To this end, we use} $\SI{1}{\percent}$ density oscillations with frequency $\SI{320}{\hertz}$ at $3.5\,n_\mathrm{sat}$ for two different temperatures, $T=\SI{5}{\kilo\eV}$ and $T=\SI{0.5}{\mega\eV}$. {At this density, we have $Y_S\simeq0.094$ and $Y_Q\simeq0.14$.} %

In Figure \ref{fig:individualtracers}, we show the resulting evolution of the deviations from chemical equilibrium $\mu_X/T$ and the pressure correction $\Pi / P_\mathrm{eq}$. For the case of $T=\SI{0.5}{\mega\eV}$, both deviations remain subthermal ($\mu_X / T < 1$). %
However, because strangeness remains near equilibrium, the bulk scalar is dominated by contributions from the deviation from $\beta$-equilibrium. Furthermore, the reaction rates of nucleonic Urca are too low to provide a significant change in $Y_e$ during the oscillation; thus, the bulk scalar oscillates almost entirely in-phase with the density oscillations and the bulk viscous damping is small.
For the case of lower temperature, $T=\SI{5}{\kilo\eV}$, $Y_e$ similarly remains frozen over the oscillation cycle, which this time yields suprathermal ($\mu_X / T > 1$) deviations. Due to the slower strangeness equilibration at this temperature, the strangeness chemical potential also becomes suprathermal, and obtains a small phase shift relative to the higher-temperature case. This deviation induces a large out-of-phase component in the bulk scalar and thus large bulk viscosity.

The evolution can also be understood from the reaction rates, which we show in Figure \ref{fig:rates}. In general, the magnitudes of the rates follow the chemical potential deviations that they affect; the non-leptonic rates follow $\mu_S$, the nucleonic Urca rates follow $\mu_n - \mu_p - \mu_e$, and the $\Lambda$ hyperon Urca is sensitive to both deviations. While the Urca rates increase with temperature, the non-leptonic rates instead saturate at both temperatures, reaching an equilibrium $\Gamma_\Lambda \simeq \dot n_\Lambda$ during the oscillation. However, for the lower temperature, these reaction rates are reached only at a suprathermal deviation in $\mu_S$, with the phase shift between the two cases arising from the timescale to reach the $\mu_S$ at which the rates saturate. This is also coherent with the flattened oscillation structure of $\mu_S$ for the $T=\SI{5}{\kilo\eV}$ case: since the required reaction rate oscillates sinusoidally and $\Gamma_\Lambda \sim -\mu_S^3$ in the suprathermal regime, the resulting relative variation in $\mu_S$ is smaller.

To characterize the regimes these oscillations probe, we find that in the higher temperature case, $\mu_S$ is small enough to be considered in equilibrium. This is also reflected by the fact that $\Pi = 0$ coincides with $\mu_\beta = 0$. Therefore, the behavior of $\Pi$ is well-described by the Israel-Stewart evolution {for single component equilibration (the limit $\lambda, \chi \rightarrow 0$ in Equation \eqref{eqn:burgers})}; with a relaxation time $\tau_\Pi \gg \omega^{-1}$ where $\omega$ is the angular frequency of the oscillation, $\Pi$ obtains a dominant in-phase response with the density oscillations. 

For the lower temperature case, we instead note that $\Pi = 0$ does not coincide exactly with $\mu_S = 0$, because both equilibrium deviations shift the pressure. This is a characteristic feature of multiple relaxing affinities in a fluid \citep{gavassinoRelativisticBulkViscous2023}. Instead, we find that with both $\mu_\beta /T \gg 1$ and $\mu_S/T \gg 1$, we are probing a far-from-equilibrium regime \citep{gavassinoRelativisticBulkRheology2024,Yang:2023ogo}, in which the linear response required for Burgers-type evolution does not hold \citep{gavassinoRelativisticBulkViscous2023}. {Such a far-from-equilibrium regime is also present in the immediate post-merger phase \citep{Most:2022yhe}.}

\begin{figure*}
    \centering
    \includegraphics[width=\linewidth]{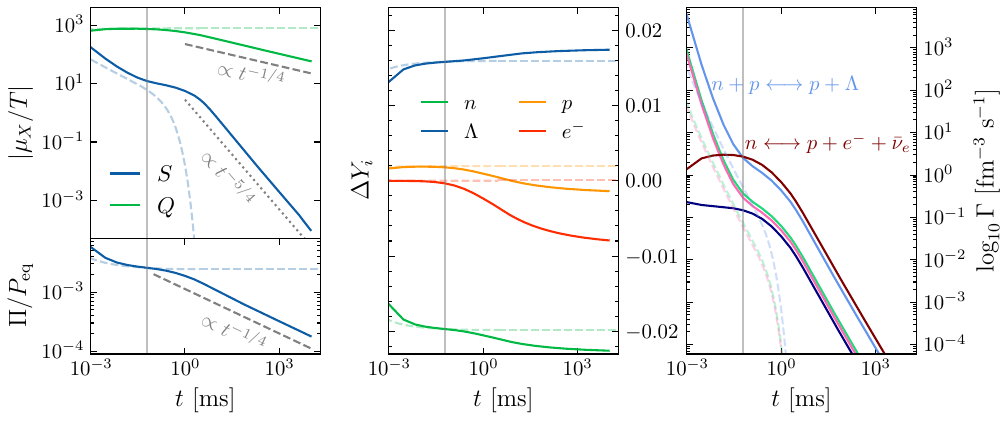}
    \caption{Decay of a suprathermal initial deviation $\mu_S/T = 10^3$, {for the case of our full set of reactions (solid lines) and the case of only nonleptonic rates (dashed lines). For the full set of reactions, the system gives} rise to polynomial decay due to coupling between the suprathermal nucleonic Urca rates and subthermal non-leptonic rates. {\textit{(Left)} Evolution of the chemical potential deviations and bulk scalar. \textit{(Middle)} Change in particle fractions, $Y_i$, relative to their initial values. \textit{(Right)} Evolution of the reaction rates, where the corresponding reaction to each color is given in the legend of Figure \ref{fig:rates}.}
    } %
    \label{fig:coupling}
\end{figure*}

To demonstrate the altered dynamics in this regime, we evolve the network from an initial state with \mbox{$\mu_\beta = 0,\, \mu_S/T = 10^3$} at $T=\SI{10}{\kilo\eV}$ and a constant density $n_B = 3.5\,n_\mathrm{sat}${, which we show in Figure \ref{fig:coupling}.} In this evolution, we find that the quick suprathermal $\mu_S$ equilibration drives $\mu_\beta$ far-from-equilibrium. In the subsequent decay, a balance between the neutron and $\Lambda$ production rates sets in, allowing $\mu_S$ to decay with polynomial time dependence on a timescale of tens of seconds. 

This time dependence can be understood on the basis of the reactions (Figure \ref{fig:coupling}): the initial large production of $\Lambda$ baryons depletes neutrons via the non-leptonic rates \eqref{reac:non-leptonic}}, which drives $\mu_\beta$ far-from-equilibrium while the charge fraction remains approximately constant, i.e., protons are not yet depleted.\footnote{{Due to the presence of other hyperons, specifically $\Sigma^-$'s, the electrons remain slightly below the proton populations.}} As $\mu_S$ transitions to the subthermal regime, this depletion of neutrons slows. Instead, the evolution becomes dominated by the far-from-equilibrium electron capture on protons, which now starts to convert protons into neutrons, counteracting the neutron depletion, which in turn can be further converted into $\Lambda$ baryons. This increases $\mu_S$ {at a rate $
-\frac{\Gamma_n}{n_B}\left.\frac{\partial \mu_S}{\partial Y_Q}\right|_{Y_S} > 0$}. 
{Thus, the total rate of change of $\mu_S$ becomes 
\begin{equation}
    \label{eq:muSchange}
    \partial_t\mu_S= -\frac{\Gamma_S}{n_B}\left.\frac{\partial \mu_S}{\partial Y_S}\right|_{Y_Q}-\frac{\Gamma_n}{n_B}\left.\frac{\partial \mu_S}{\partial Y_Q}\right|_{Y_S}.
\end{equation}}
This effectively yields an approximate balancing of the strangeness changing and neutron production rates, $|\Gamma_S| \sim |\Gamma_n|$. With suprathermal $\beta$-rates, we have $\partial_t {\mu_\beta} \sim \Gamma_{n} \sim -\mu_\beta^5$. This gives a relaxation $\mu_\beta\sim t^{-1/4}$, giving the $\mu_S$ decay as $\mu_S \sim |\Gamma_S| \sim |\Gamma_n| \sim t^{-5/4}$.\footnote{{This can also be understood by solving $y' = -ay(x) + bx^{-5/4}$, corresponding to Equation \eqref{eq:muSchange}. For large $x$, the solution approaches $ay(x) \simeq bx^{-5/4}$, giving the expected rate balance.}} This demonstrates a coupling between the rates causing significantly slower strangeness equilibration {that is limited by the $\beta$-equilibration}, and explains the observed behavior in Figure \ref{fig:coupling}, including the sustained bulk viscous damping despite the intrinsic mismatch between strangeness and $\beta$-equilibration timescales.\\

For comparison, we also show in Figure \ref{fig:coupling} the evolution of the system when only the nonleptonic strangeness-changing rates are included. In this case, we still find that $\mu_\beta$ is driven far-from-equilibrium, but $\mu_S$ decays on a timescale of milliseconds due to the absence of the coupling described above, {lacking the net change in $Y_e$ that is caused by electron capture in the coupled case.}

\subsection{Burgers dynamics}
\label{subsec:burgers}
To connect our reaction network calculations to the Burgers fluid description of multicomponent fluids in the linear response regime \citep{gavassinoRelativisticBulkViscous2023, hernandezBurgersEquationBulk2025}, we compare the evolution of Equation \eqref{eqn:burgers} to reaction network evolutions. We again consider oscillations at $n_B = 3.5\,n_\mathrm{sat}$ for initial temperatures $T = \SI{5}{\kilo\eV}$ and $T = \SI{0.5}{\mega\eV}$. We now include frequencies of $\SI{160}{\hertz}$, $\SI{320}{\hertz}$, $\SI{480}{\hertz}$, $\SI{640}{\hertz}$, and $\SI{960}{\hertz}$, with small-amplitude oscillations of $\SI{0.01}{\percent}$ and large-amplitude oscillations of $\SI{1}{\percent}$.

For a harmonic oscillation, the bulk scalar can be split into a dissipative and reactive part as
\begin{equation}
    \Pi(t) = -\zeta_\mathrm{eff}(\omega)\theta(t) + \chi_\mathrm{eff}(\omega)\dot\theta(t),
\end{equation}
where $\zeta_\mathrm{eff}$ is the effective bulk viscosity giving rise to dissipation, and $\chi_\mathrm{eff}$ parametrizes the reactive response of $\Pi$. From the reaction network, we extract the effective bulk viscosity $\zeta_\mathrm{eff}(\omega)$ as $\zeta_\mathrm{eff} = -\langle \Pi \theta \rangle / \langle \theta^2 \rangle$, which is the least-squares estimator of the effective {AC bulk viscosity (see also \citet{Yang:2023ogo})} in Equation \eqref{eq:NSlimit}. We similarly extract the effective reactive parameter as $\chi_\mathrm{eff} = -\langle \Pi \dot\theta \rangle / \langle \dot\theta^2 \rangle$. By fitting the reaction network evolution to Equation \eqref{eqn:burgers}, we can also compute $\zeta_\mathrm{eff}$ as 
\begin{equation}
    \zeta_\mathrm{eff} = \frac{\zeta (1 - \lambda \omega^2) + \chi \tau_\Pi\omega^2}{(1 - \lambda \omega^2)^2 + \omega^2\tau_\Pi^2},
\end{equation}
and the reactive response parameter as
\begin{equation}
    \chi_\mathrm{eff} = \frac{\zeta \tau_\Pi - \chi(1 - \lambda\omega^2)}{(1 - \lambda\omega^2)^2 + \omega^2\tau_\Pi^2}.
\end{equation}
In the frozen limit ($\lambda, \chi \to 0$, $\tau_\Pi\omega \to \infty$), which we found for the large-amplitude $T=\SI{0.5}{\mega\eV}$ oscillations in Section \ref{subsec:reactiondynamics}, we note that $\omega\chi_\mathrm{eff} \to \zeta / (\tau_\Pi\omega)$. In the low-frequency limit ($\omega \to 0$), we similarly have $\zeta_\mathrm{eff} \to \zeta$.

\begin{figure}
    \centering
    \includegraphics[width=1.0\linewidth]{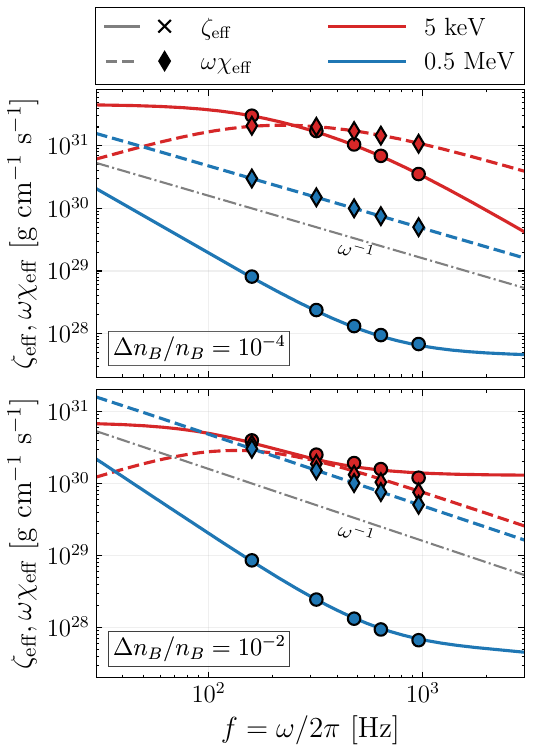}
    \caption{Comparison of the dissipative, $\zeta_\mathrm{eff}$, and reactive, $\omega\chi_\mathrm{eff}$, components of the bulk scalar for sinusoidal $\SI{0.01}{\percent}$ (upper) and $\SI{1}{\percent}$ (lower) frequency oscillations. Solid and dashed lines show the predictions of $\zeta_\mathrm{eff}$ and $\omega\chi_\mathrm{eff}$, respectively, from the fitted Burgers evolution. Circles and diamonds mark the reaction network results of $\zeta_\mathrm{eff}$ and $\omega\chi_\mathrm{eff}$, respectively. Results for $\SI{5}{\kilo\eV}$ are shown in red, and $\SI{0.5}{\mega\eV}$ in blue.}
    \label{fig:Burgers}
\end{figure}

In Figure \ref{fig:Burgers}, we compare the values of $\zeta_\mathrm{eff}$ and $\omega\chi_\mathrm{eff}$ extracted from the reaction network evolution and from the fit to Equation \eqref{eqn:burgers}. We find that for small-amplitude oscillations, the Burgers-type evolution agrees well with the values extracted from the reaction network, indicating that these oscillations are well-described as Burgers dynamics. For the larger $\SI{1}{\percent}$ oscillations, the errors for the $\SI{5}{\kilo\eV}$ case grow as the evolution probes far-from-equilibrium regimes in which the linear response assumption underlying the Burgers description is invalid.  We also confirm for both oscillation amplitudes that the $T = \SI{0.5}{\mega\eV}$ evolution is well-described by the frozen limit of Equation \eqref{eqn:burgers}, with $\omega\chi_\mathrm{eff} \propto \omega^{-1}$ across the entire range of frequencies shown, for both amplitudes.

We can also compare the parameters of the fit to their expected values. For the $\SI{0.01}{\percent}$ oscillations with $T=\SI{0.5}{\mega\eV}$, we find the fitted relaxation parameters $\lambda \simeq \SI{2.1e-7}{\second\squared}$ and $\tau_\Pi \simeq \SI{40.4}{\milli\second}$. By computing the two relaxation times $\tau_1, \tau_2$ for the diagonalized form of Equation \ref{eqn:burgers}, corresponding to splitting the total response into two independent Israel-Stewart relaxations (see Refs. \citet{gavassinoRelativisticBulkViscous2023, hernandezBurgersEquationBulk2025}), we obtain $\tau_1 \simeq \SI{40.4}{\milli\second}$ and $\tau_2 \simeq \SI{0.0052}{\milli\second}$. The value of $\tau_1$ agrees well with the expected timescale of equilibration at this temperature: the Israel-Stewart $\beta$-relaxation time  
\begin{equation}
    \tau^{-1} = \left.\frac{\partial \delta \mu_\beta}{\partial Y_e}\right|_{\varepsilon, n_B, Y_e^\mathrm{eq}, \mu_S = 0}\frac{\lambda_0}{n_B}
\end{equation}
(see \citet{Most:2022yhe}) is $\tau \simeq \SI{38.3}{\milli\second}$, and by computing the eigenmodes from the reaction matrix (see Equation\ (22) of \citet{gavassinoRelativisticBulkViscous2023}) we find $\tau_1\simeq \SI{42.5}{\milli\second}$. By contrast, the fitted value of $\tau_2$ is instead two orders of magnitude larger than expected from the eigenmode decomposition.
Conversely, for the $T=\SI{5}{\kilo\eV}$ case, the expected slow eigenmode has $\tau_1 \simeq \SI{4.25e6}{\second}$, which is not captured by our fitting as it far exceeds the timescale of our evolutions.
Thus, while the fitting exhibits some intrinsic limitations in parameter estimation, we conclude that the reaction network evolution can probe the Burgers evolution of the system and extend beyond it for large-amplitude oscillations.

\subsection{Bulk viscosity}
\label{subsec:bulkviscosity}

After analyzing the rate dynamics which determine the far-from-equilibrium response of the matter, we now move on to computing the bulk viscosity. We extract it using Equation \eqref{eq:NSlimit} as a function of temperature for small sinusoidal oscillations about equilibrium at the $\SI{1.95}{\Msun}$ g-mode frequency of $\SI{320}{\hertz}$. 
We compare the effective bulk viscosity with strangeness in and out-of-equilibrium, and the results are shown in Figure \ref{fig:linearresponsebulk}. We also show, with fainter lines, the results of $\SI{0.5}{\percent}$ density oscillations. Generally, if the density is large enough for strangeness to emerge, the faster strangeness equilibration rates result in significant bulk viscosity at much lower temperatures than for the strangeness-equilibrated case. Due to its weaker temperature dependence, the strangeness-driven bulk viscosity is also relevant for a much wider range of temperatures that are expected during inspiral \citep{Arras:2018fxj, ghoshTidalHeatingDirect2024}. At higher temperatures, strangeness becomes equilibrated ($\mu_S \approx 0$), while bulk viscosity emerges due to direct Urca reactions, or modified Urca reactions below the direct Urca threshold ($n_B < \SI{0.31}{\per\femto\meter\cubed}$ for the CMF EOS). The resonant peaks we find in the bulk viscosity that are seen for all three equilibration channels are expected in the near-equilibrium response and are in general agreement with previous calculations \citep{alfordDampingDensityOscillations2019, alfordStrangenesschangingRatesHyperonic2021}. 

For the case of $\SI{0.5}{\percent}$ oscillations, we find that the bulk viscosity approaches a constant value before the resonant peak as the temperature is lowered. This is consistent with high-amplitude oscillations that probe the suprathermal regime, as has previously been studied for $\beta$-equilibration \citep{alfordLargeAmplitudeBehavior2010}. This also indicates that for stars with hyperonic cores, the bulk viscous heating could be significant for a wide range of temperatures if core compression during resonance is sufficiently large.

\begin{figure}
    \centering
    \includegraphics[width=\linewidth]{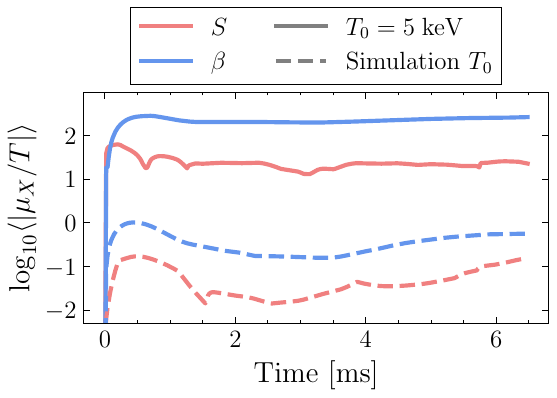}
    \caption{{Density-averaged chemical potential deviations, $\mu_X/T$, in the evolutions of the fluid elements for the initial temperature set to $T_0=\SI{5}{\kilo\eV}$ (solid lines) and taken from the simulation (dashed lines). {Here, $\beta$ and $S$ refer to the out-of-equilibrium chemical potentials for $\beta$- and strangeness processes, respectively.}}}
    \label{fig:simulation}
\end{figure}

Finally, we evaluate the bulk viscosity in the fluid elements of a binary neutron star inspiral simulation. We evolve the composition with and without strangeness out-of-equilibrium, and extract the {density-averaged} bulk viscosity and chemical deviations (see Figure \ref{fig:simulation}). We find that, below the hyperon onset, the average bulk viscosity is $\sim \SI{2e28}{\gram\per\centi\meter\per\second}$, caused by large far-from-equilibrium deviations from $\beta$-equilibrium with $\langle |\mu_\beta / T |\rangle \sim 420$. Above the hyperon onset, the average bulk viscosity with strangeness out-of-equilibrium is instead $\sim \SI{2e29}{\gram\per\centi\meter\per\second}$, with far-from-equilibrium regimes probed for both strangeness and $\beta$-equilibrium, with $\langle |\mu_S / T |\rangle \sim 80$ and $\langle |\mu_\beta / T |\rangle \sim 260$. For the case with equilibrated strangeness, the bulk viscosity above hyperon onset is instead {two orders of magnitude smaller at } $\sim \SI{8e27}{\gram\per\centi\meter\per\second}$, with $\langle |\mu_\beta / T |\rangle \sim 250$. 

We thus find that the inspiral fluid elements we analyze probe the far-from-equilibrium dynamics discussed in Section \ref{subsec:reactiondynamics}, and the hyperonic core exhibits a significantly increased bulk viscosity when strangeness is pushed out-of-equilibrium. However, we note that the bulk viscosity above the hyperon threshold is lower than in Figure \ref{fig:linearresponsebulk}. Since bulk viscosity generally decreases with oscillation frequency, we attribute this lower bulk viscosity to high-frequency oscillations caused by errors in the initial data \citep{kuanErrorBudgetBinary2025}. 
As an additional validation, we manually filter out these high frequency oscillations from the fluid trajectories. We then find that in either case a far-from-equilibrium regime persists above the hyperon threshold, with $\langle |\mu_S / T|\rangle \sim 25$ and $\langle |\mu_\beta / T|\rangle \sim 220$. Similarly, below the hyperon threshold, we still find a large value of  $\langle |\mu_\beta / T|\rangle \sim 390$. 

Finally, it is instructive to compare these physically expected temperatures \citep{Arras:2018fxj} to the ones probed in the simulation (see Section \ref{subsec:inspiralsimulation}). In line with \citet{gittinsProblematicSystematicsNeutronstar2025} our simulation has substantial artificial heating of the stars, leading to  $\langle |\mu_S / T|\rangle \sim 0.06$ and $\langle |\mu_\beta / T|\rangle \sim 0.4$. Below the hyperon threshold, we similarly get a much smaller deviation of $\langle |\mu_\beta / T|\rangle \sim 1.62$. This demonstrates that the high temperatures induced by numerical surface heating \citep{gittinsProblematicSystematicsNeutronstar2025} strongly suppress any far-from-equilibrium dynamics, which according to our analysis should be present. Consequently, current numerical relativity simulations fundamentally probe both the wrong thermodynamic and chemical reaction regime.
More work with accurate inspiral simulations will be required to address this point fully.

\section{Conclusion}
In this work, we have presented the first {fully coupled} reaction network which can probe the coupled far-from-equilibrium dynamics of strangeness and $\beta$-equilibration in neutron star mergers {on top of numerical relativity backgrounds}. By dynamically evolving the composition of the matter, we evaluate the effective bulk viscosity of matter with and without the effects of strangeness equilibration. We also demonstrate, by tracking the reaction rates in the evolution, that the reaction network can probe far-from-equilibrium regimes in which the strangeness and $\beta$ relaxations can become coupled. Finally, we assess the role of strangeness equilibration on binary neutron star inspirals, and find that, for sufficiently massive stars, 
the inspiral can probe far-from-equilibrium regimes from both strangeness and $\beta$-equilibrium, and the resulting dissipation could substantially affect the inspiral evolution, leading to phase shifts, $\Delta \Psi \simeq 0.14\,\rm rad$.

Our results indicate that, at the low temperatures during inspiral, hyperonic bulk viscosity could be dominant for the dynamical properties of massive neutron stars. In particular, it is possible that substantial heating and orbital phase shifts could be associated with both continuous and resonant hyperonic-induced dissipation. However, a full understanding of the extent of these effects will ultimately require self-consistent inclusion of strangeness equilibration, $\beta$ equilibration, and associated neutrino energy losses, as well as control of artificial shock heating in inspiral simulations \citep{gittinsProblematicSystematicsNeutronstar2025}. 
Such an understanding could potentially enable dynamical probes of strangeness to rule on its existence in neutron star matter.

Finally, many other channels may alter or induce additional bulk viscosity. The presence of muons can increase the Urca threshold density and widen the temperature range of Urca-driven bulk viscosity \citep{alfordLeptonicContributionBulk2010, alfordBulkViscosityRelativistic2022, alfordBulkViscosityUrca2023}. Moreover, the influence of pions and pion condensates \citep{pajkosInfluenceMuonsPions2025, harrisBulkViscosityNuclear2025} remains to be understood. 
{Additionally, other hadrons (like $\Delta$ baryons) or hyperons could be included, which would significantly complicate the coupled rate equations.}
Finally, if a first-order deconfinement phase transition occurs in the cores of massive neutron stars, this could induce an interface oscillation mode with stronger tidal excitation than the g-mode \citep{counsellInterfaceModesInspiralling2025} and open the possibility of bulk viscosity from weak equilibration of quarks \citep{alfordBulkViscosityTwocolor2024, hernandezBurgersEquationBulk2025} or phase conversion dissipation \citep{alfordPhaseConversionDissipation2015}.

\begin{acknowledgments}
The authors are grateful for insightful discussions with Abhishek Hegade, Jorge Noronha, Yumu Yang, and Nicolas Yunes.
The authors acknowledge support by the U.S. National Science Foundation under the MUSES collaboration through grant OAC-2103680.
{MS acknowledges support through Caltech's SURF and VURP programs.}
ERM, MS and JW acknowledge support by the U.S. National Science Foundation under Grant Nos. PHY-2541792, PHY-2309210, and AST-2510568. ERM is also supported by a Research Fellowship from the Sloan Foundation and a William H. Hurt Scholarship at the California Institute of Technology.
J.N.H. acknowledges support from the US-DOE Nuclear Science
Grant No. DE-SC0023861.
A.H.~acknowledges financial support by the UKRI under the Horizon Europe Guarantee project EP/Z000939/1. V.D. acknowledges support from the U.S. Department of Energy, Office of Science, Nuclear Physics program under Grant
DE-SC0024700 and from the National Science Foundation under grant NP3M PHY2116686.
The simulations were performed on Delta at the National Center for Supercomputing Applications (NCSA) through allocation PHY210074 from the Advanced Cyberinfrastructure Coordination Ecosystem: Services \& Support (ACCESS) program, which is supported by National Science Foundation grants \#2138259, \#2138286, \#2138307, \#2137603, and \#2138296. Support also comes from the Resnick High Performance Computing Center, a facility supported by the Resnick Sustainability Institute at the California Institute of Technology.
\end{acknowledgments}

\appendix

\section{Multi-component chemical equilibration}
\label{app:burgers}

Following \citet{gavassinoRelativisticBulkViscous2023}, we can write the general form of multi-component chemical relaxation. 
We assume a perfect fluid, $T^{\mu\nu} = (\varepsilon+P)\,u^\mu u^\nu + P g^{\mu\nu}$, with
$P = P(\mathfrak{s},v,Y_a)$, for species fraction, $Y_a$, where $u^\mu$ is the four velocity, $\varepsilon$ the energy density, $P$ the pressure, $\mathfrak{s}$ is the specific entropy per baryon, and $g^{\mu\nu}$ the inverse four-metric. The first law of thermodynamics then states $d\epsilon = T d\mathfrak{s} - P dv - \mathbb{A}^a dY_a$, {where $\epsilon$} is the specific internal energy 
, $v = 1/n_B$, for baryon density $n_B$.
The fluid has both baryon conservation, $\nabla_\mu (n_B u^\mu)=0$, and energy momentum conservation, $\nabla_\mu T^{\mu\nu}=0$.
Thermodynamics and energy conservation imply
\begin{equation}\label{eqn:app_entropy}
  T\dot{\mathfrak{s}} = \mathbb{A}^a \dot{Y}_a
    = \frac{1}{n_B}\,\mathbb{A}^a \Gamma_a =: \frac{1}{n_B}\,\mathcal{Q}\,,
\end{equation}
where we have used the rate equation per species $\dot{Y}_a = v\Gamma_a$, and introduced the shorthand {$\dot{Z} = u^\mu \nabla_\mu Z$ for the comoving time derivative of a variable $Z$.}  Here $\mathcal{Q} = \mathbb{A}^a \Gamma_a$ is the heating rate of the reactions.
We further write,
\begin{equation}
  \dot{Y}_a
  = \left.\frac{\partial Y_a}{\partial v}\right|_{\mathfrak{s},\mathbb{A}}\!\dot{v}
  + \left.\frac{\partial Y_a}{\partial \mathfrak{s}}\right|_{v,\mathbb{A}}\!\dot{\mathfrak{s}}
  + \left.\frac{\partial Y_a}{\partial \mathbb{A}^b}\right|_{\mathfrak{s},v}\!\dot{\mathbb{A}}^b\,,
  \label{eq:app_chain}
\end{equation}

and solve this equation for $\dot{\mathbb{A}}_a$,
\begin{equation}
  v\mathcal{R}_a
  = -\mathcal{C}_a\,(v\theta)
  + T\mathcal{S}_a\,\frac{v\mathcal{Q}}{T}
  - \frac{\mathcal{B}_{ab}}{n}\,\dot{\mathbb{A}}^b\,,
  \label{eq:app_collect}
\end{equation}
which then provides the evolution equation for the chemical potentials.

\section{Rate equations}
\label{sec:rateequations}
Here, we explicitly write the evolution equations of the particle fractions for neutrons, protons, electrons, and $\Lambda$ hyperons in terms of their contributing reactions.  Including {the $npe$ Urca reactions and} all reactions shown in Equations (\ref{reac:strangeUrca}--\ref{reac:non-leptonic}), we can write down the rate equations for the particle fractions, $Y_i$, of each species if they were allowed to vary independently of each other:
\begin{align}
\dot Y_\Lambda ={}&
n_B\sigma_{nn}\left[\frac{Y_\Lambda^{\rm eq}}{Y_n^{\rm eq}}Y_n^2-Y_\Lambda Y_n\right]
+n_B{\sigma_{np}}\left[\frac{Y_\Lambda^{\rm eq}}{Y_n^{\rm eq}}Y_n Y_p-Y_\Lambda Y_p\right]
+n_B{\sigma_{n\Lambda}}\left[\frac{Y_\Lambda^{\rm eq}}{Y_n^{\rm eq}}Y_n Y_\Lambda-Y_\Lambda^2\right]
\nonumber\\
&+\Gamma_{\Lambda,\beta}\left[\frac{Y_\Lambda^{\rm eq}}{Y_p^{\rm eq}Y_e^{\rm eq}}Y_pY_e-Y_\Lambda\right]
+n_B{\sigma_{pe,\Lambda\nu}}\left[Y_pY_e-\frac{Y_p^{\rm eq}Y_e^{\rm eq}}{Y_\Lambda^{\rm eq}}Y_\Lambda\right],
\label{eq:nLambda_reduced}
\\[0.5em]
\dot Y_n ={}&
n_B{\sigma_{nn}}\left[Y_\Lambda Y_n-\frac{Y_\Lambda^{\rm eq}}{Y_n^{\rm eq}}Y_n^2\right]
+n_B{\sigma_{np}}\left[Y_\Lambda Y_p-\frac{Y_\Lambda^{\rm eq}}{Y_n^{\rm eq}}Y_n Y_p\right]
+{n_B}{\sigma_{n\Lambda}}\left[Y_\Lambda^2-\frac{Y_\Lambda^{\rm eq}}{Y_n^{\rm eq}}Y_n Y_\Lambda\right]
\nonumber\\
&+\Gamma_n\left[\frac{Y_n^{\rm eq}}{Y_p^{\rm eq}Y_e^{\rm eq}}Y_pY_e-Y_n\right]
+n_B\sigma_{pe^-}\left[Y_pY_e-\frac{Y_p^{\rm eq}Y_e^{\rm eq}}{Y_n^{\rm eq}}Y_n\right],
\label{eq:nn_reduced}
\\[0.5em]
\dot Y_p ={}&
\Gamma_n\left[Y_n-\frac{Y_n^{\rm eq}}{Y_p^{\rm eq}Y_e^{\rm eq}}Y_pY_e\right]
+n_B\sigma_{pe^-}\left[\frac{Y_p^{\rm eq}Y_e^{\rm eq}}{Y_n^{\rm eq}}Y_n-Y_pY_e\right]
+\Gamma_{\Lambda,\beta}\left[Y_\Lambda-\frac{Y_\Lambda^{\rm eq}}{Y_p^{\rm eq}Y_e^{\rm eq}}Y_pY_e\right] \nonumber \\ 
&+n_B\sigma_{pe^-,\Lambda\nu}\left[\frac{Y_p^{\rm eq}Y_e^{\rm eq}}{Y_\Lambda^{\rm eq}}Y_\Lambda-Y_pY_e\right],
\label{eq:np_reduced}
\\[0.5em]
\dot Y_e ={}&
\Gamma_n\left[Y_n-\frac{Y_n^{\rm eq}}{Y_p^{\rm eq}Y_e^{\rm eq}}Y_pn_e\right]
+n_B\sigma_{pe^-}\left[\frac{Y_p^{\rm eq}Y_e^{\rm eq}}{Y_n^{\rm eq}}Y_n-Y_pY_e\right]
+\Gamma_{\Lambda,\beta}\left[Y_\Lambda-\frac{Y_\Lambda^{\rm eq}}{Y_p^{\rm eq}Y_e^{\rm eq}}Y_pY_e\right] \nonumber \\ 
&+n_B\sigma_{pe^-,\Lambda\nu}\left[\frac{Y_p^{\rm eq}Y_e^{\rm eq}}{Y_\Lambda^{\rm eq}}Y_\Lambda-Y_pY_e\right].
\label{eq:ne_reduced}
\end{align}
where the equilibrium values, $Y_i^{\rm eq} {= n_i^{\rm eq}/n_B}$, are obtained from an underlying EOS {, $\Gamma_i$ denotes a decay width, and $\sigma_{ij}$ can similarly be interpreted as the number of interactions per unit time. }

Using the generic rate equations above for $np\Lambda e$ matter, we identify the two terms in the gain--loss expression for each Urca channel with the corresponding decay and capture rates. For example,
\begin{equation}
\Gamma_n n_n\equiv\Gamma_{\mathrm{dU},n,\mathrm{decay}},
\qquad
\Gamma_n\frac{n_n^{\rm eq}}{n_p^{\rm eq}n_e^{\rm eq}}n_pn_e
\equiv\Gamma_{\mathrm{dU},n,e^-\mathrm{\ capture}},
\end{equation}
where the relative normalization of the reverse process is fixed by detailed balance, so that
\begin{equation}
\Gamma_n\left[n_n-\frac{n_n^{\rm eq}}{n_p^{\rm eq}n_e^{\rm eq}}n_pn_e\right]
=\Gamma_{\mathrm{dU},n,\mathrm{decay}}
-\Gamma_{\mathrm{dU},n,e^-\mathrm{\ capture}}
\equiv\Gamma_{\mathrm{dU},n}.
\end{equation}
The same identification is made for the hyperonic Urca and non-leptonic gain--loss terms.
Then, in the co-moving frame of the fluid, Equation \eqref{eq:fractionevolution} becomes
\begin{align}
    &\begin{aligned}
        \label{eq:Yn_evolution}\frac{dY_n}{d\tau} &= -\frac{1}{n_B}\underbrace{\left(\Gamma_{\mathrm{dU, {n}, decay}} - \Gamma_\mathrm{dU, {n}, \text{$e^-$ capture}} \right)}_{\Gamma_\mathrm{dU, {n}}} -\frac{1}{n_B}\underbrace{\left(\Gamma_{\mathrm{mU(n), decay}} - \Gamma_\mathrm{mU(n), \text{$e^-$ capture}}\right)}_{\Gamma_{\mathrm{mU(n)}}}
    \\ &- \frac{1}{n_B}\underbrace{\left(\Gamma_{\mathrm{mU(p), decay}} - \Gamma_\mathrm{mU(p), \text{$e^-$ capture}}\right)}_{\Gamma_{\mathrm{mU(p)}}}
    - \frac{1}{n_B}\Gamma_{nn\leftrightarrow n\Lambda} - \frac{1}{n_B}\Gamma_{np\leftrightarrow \Lambda p} - \frac{1}{n_B}\Gamma_{n\Lambda\leftrightarrow \Lambda \Lambda} + \frac{1}{n_B}\Gamma_{\mathrm{strong, n}}, \\
    \end{aligned} \\
    &\begin{aligned}
    \label{eq:Ye_evolution}\frac{dY_e}{d\tau} &= \frac{1}{n_B}\underbrace{\left(\Gamma_{\mathrm{dU, {n}, decay}} - \Gamma_\mathrm{dU, {n}, \text{$e^-$ capture}} \right)}_{\Gamma_\mathrm{dU, {n}}} +\frac{1}{n_B}\underbrace{\left(\Gamma_{\mathrm{mU(n), decay}} - \Gamma_\mathrm{mU(n), \text{$e^-$ capture}}\right)}_{\Gamma_{\mathrm{mU(n)}}}
    \\ &+ \frac{1}{n_B}\underbrace{\left(\Gamma_{\mathrm{mU(p), decay}} - \Gamma_\mathrm{mU(p), \text{$e^-$ capture}}\right)}_{\Gamma_{\mathrm{mU(p)}}} + \frac{1}{n_B}\underbrace{\left(\Gamma_{\mathrm{dU, {\Lambda}, decay}} - \Gamma_\mathrm{dU, {\Lambda}, \text{$e^-$ capture}} \right)}_{\Gamma_\mathrm{dU, {\Lambda}}}, \\
    \end{aligned}\\
    &\begin{aligned}
    \label{eq:Yp_evolution}\frac{dY_p}{d\tau} &= \frac{dY_e}{d\tau} + \frac{1}{n_B}\Gamma_{\mathrm{strong, p}},
    \end{aligned}\\
    &\begin{aligned}
    \label{eq:YLambda_evolution}\frac{d Y_\Lambda}{d\tau} &= - \frac{1}{n_B}\underbrace{\left(\Gamma_{\mathrm{dU, {\Lambda}, decay}} - \Gamma_\mathrm{dU, {\Lambda}, \text{$e^-$ capture}} \right)}_{\Gamma_\mathrm{dU, {\Lambda}}} + \frac{1}{n_B}\Gamma_{nn\leftrightarrow n\Lambda} + \frac{1}{n_B}\Gamma_{np\leftrightarrow \Lambda p} + \frac{1}{n_B}\Gamma_{n\Lambda\leftrightarrow \Lambda \Lambda} + \frac{1}{n_B}\Gamma_{\mathrm{strong, \Lambda}},
    \end{aligned}
\end{align}
where $\Gamma_{\mathrm{dU, n}}$ and $\Gamma_{\mathrm{dU, \Lambda}}$ are the volumetric neutron and $\Lambda$ direct Urca rates, $\Gamma_{\mathrm{mU(n)}}$ and $\Gamma_{\mathrm{mU(p)}}$ are the volumetric modified Urca rates with neutron and proton spectators, and $\Gamma_{nn\leftrightarrow n\Lambda}$, $\Gamma_{np\leftrightarrow \Lambda p}$, and $\Gamma_{n\Lambda\leftrightarrow \Lambda \Lambda}$ are the volumetric non-leptonic hyperon reactions of Equation \eqref{reac:non-leptonic}. The rates of direct Urca, modified Urca, and non-leptonic strange reactions are given in Appendices \ref{sec:dUrca}, \ref{sec:mUrca}, and \ref{sec:non-leptonic}, respectively. The rate contributions $\Gamma_{\mathrm{strong, n}}$, $\Gamma_{\mathrm{strong, p}}$, and $\Gamma_{\mathrm{strong, \Lambda}}$ account for the redistribution of baryons through strong interactions.

We do not solve for these rates directly; instead, they are fixed (by instantaneous equilibration) at the level of the CMF EOS. 
What that means is that we enforce  charge neutrality via 
\begin{equation}\label{eqn:electric_cons_constraint}
n_p\sim n_e=n_BY_e,\qquad \dot{n}_e \sim  \dot{n}_p
\end{equation}
and baryon conservation via
\begin{eqnarray}\label{eqn:baryon_cons_constraint}
    \dot{n}_B&=&\dot{n}_n+\dot{n}_p+\dot{n}_{\Lambda} %
\end{eqnarray}
that together provides the constraint $Y_n=1-Y_e-Y_S$, and $Y_p = Y_e$. 
Note that Equations \ (\ref{eqn:electric_cons_constraint},\ref{eqn:baryon_cons_constraint}) are only approximate because the CMF EOS that we used does have a sub-dominant contribution from $\Sigma^-$ and $\Xi$'s as well. Future work will explore their contribution.
Our EOS depends on four variables $(n_B,T,Y_e,Y_S)$, so that we only track the net electron (charge) fraction and the strangeness fraction.
Consequently, our network then evolves $Y_e$ via Equation \eqref{eq:Ye_evolution} and $Y_S$ as follows,
\begin{align}
    {\frac{d Y_S}{d\tau}} &{= - \frac{1}{n_B}\underbrace{\left(\Gamma_{\mathrm{dU, {\Lambda}, decay}} - \Gamma_\mathrm{dU, {\Lambda}, \text{$e^-$ capture}} \right)}_{\Gamma_\mathrm{dU, {\Lambda}}} + \frac{1}{n_B}\Gamma_{nn\leftrightarrow n\Lambda} + \frac{1}{n_B}\Gamma_{np\leftrightarrow \Lambda p} + \frac{1}{n_B}\Gamma_{n\Lambda\leftrightarrow \Lambda \Lambda}\,.}
\end{align}
Once the evolution is obtained, we can recover individual particle fractions via the EOS, i.e., $Y_n = Y_n(n_B,T,Y_e,Y_S)$, and so on.

\section{Reaction rates}
\label{app:reactionrates}
For the reactions considered here, we compute reaction rates $\Gamma_i$ using the Fermi surface approximation, which assumes that the matter is degenerate and that only particles close to their Fermi surfaces contribute to the reaction rate. Above $T\simeq\SI{1}{\mega\eV}$, thermal blurring invalidates this approximation and, for reactions involving neutrinos, alters the condition for $\beta$-equilibrium when neutrinos are free-streaming \citep{alford$ensuremathbeta$EquilibriumNeutronstar2018, alfordBetaEquilibriumNeutron2021}. However, for the sub-MeV temperatures considered in this work, these thermal corrections are expected to remain negligible.
\section{Direct Urca rates} \label{sec:dUrca}
For the direct Urca reactions the net reaction rate is given by \citep{Yakovlev:2000jp}
\begin{equation}
    \begin{split}
    \Gamma_{\mathrm{dU}} &=  \Gamma_{\mathrm{dU, decay}} - \Gamma_\mathrm{dU, \text{$e^-$ capture}} \\ 
    &= \frac{G^2(f^2+ 3g^2_A)}{240\pi^5}E_{Fi}^*E_{Fp}^*k_{Fe}\theta_{dU} \\ &~~~~~\times \delta\mu(17\pi^4T^4 + 10\pi^2\delta\mu^2T^2 + \delta\mu^4),
    \end{split} 
\end{equation}
where the particle $i$ is either a neutron or a $\Lambda$ hyperon, $E^*_{Fi} = \sqrt{k_{Fi}^2 + m_{i}^{*\, 2}}$ is the Fermi energy of particle $i$, $k_{Fi}$ is its Fermi momentum, and $m_i^*$ is the effective mass, which we take from the out-of-equilibrium CMF EOS. If $i$ is a neutron, $\delta\mu = \mu_n - \mu_p - \mu_e$, $f=1$, $g_A = 1.26$, and $G = G_F\cos\theta_c$, where $G_F = \SI{1.166e-11}{\per\mega\eV\squared}$ is the Fermi constant and $\theta_c = 13.04^\circ$ is the Cabibbo angle. If $i$ is a $\Lambda$ hyperon, $\delta\mu = \mu_\Lambda - \mu_p - \mu_e$, $f=-1.225$, $g_A = 0.893$, and $G = G_F\sin\theta_C$. The threshold factor
\begin{equation}
    \theta_{\mathrm{dU}} = \begin{cases}
        1 & k_{Fi} < k_{Fp} + k_{Fe},\\
        0 & \text{otherwise}
    \end{cases}
\end{equation}
ensures momentum conservation in the Fermi surface approximation and provides the direct Urca threshold criterion.

\section{Modified Urca rates}\label{sec:mUrca}

Under the same assumptions as for the direct Urca rates, the net rate of modified Urca with a neutron spectator ($N = n$) takes the form \citep{alfordBetaEquilibriumNeutron2021}
\begin{equation}
    \begin{split}
    &\Gamma_{\mathrm{mU(n)}} = \Gamma_{\mathrm{mU(n), decay}} - \Gamma_\mathrm{mU(n), \text{$e^-$ capture}} \\ 
    &= \frac{1}{5760\pi^9}G_F^2\cos^2\theta_cg_A^2f^4\frac{\left(E_{Fn}^*\right)^3E^*_{Fp}}{m_\pi^4}\frac{k^4_{Fn}k_{Fp}}{\left(k_{Fn}^2 + m_\pi^2\right)^2}\theta_n \\
    & \times \delta\mu\left(1835\pi^6T^6 + 945\pi^4\delta\mu^2T^4 + 105\pi^2\delta\mu^4T^2 + 3\delta\mu^6\right),
    \end{split} 
\end{equation}
where $\delta\mu = \mu_n - \mu_p - \mu_e$, $m_\pi$ is the pion mass, $f\approx 1$ is the nucleon-pion coupling, and 
\begin{equation}
    \theta_n = \begin{cases}
        1 & k_{Fi} < k_{Fp} + k_{Fe},\\
        1 - \dfrac{3}{8}\dfrac{(k_{Fp} + k_{Fe} - k_{Fn})^2}{k_{Fp}k_{Fe}} & \text{otherwise}.
    \end{cases}
\end{equation}
The net rate of proton-spectator ($N = p$) modified Urca is similarly \citep{alfordBetaEquilibriumNeutron2021}
\begin{equation}
    \begin{split}
    &\Gamma_{\mathrm{mU(p)}} = \Gamma_{\mathrm{mU(p), decay}} - \Gamma_\mathrm{mU(p), \text{$e^-$ capture}} \\ 
    &= \frac{1}{40320\pi^9}G_F^2\cos^2{\theta_c}g_A^2f^4\frac{\left(E_{Fp}^*\right)^3E^*_{Fn}}{m_\pi^4}\frac{(k_{Fn} - k_{Fp})^4k_{Fn}}{\left(\left(k_{Fn} - k_{Fp}\right)^2 + m_\pi^2\right)^2}\theta_p \\
    & \times \delta\mu\left(1835\pi^6T^6 + 945\pi^4\delta\mu^2T^4 + 105\pi^2\delta\mu^4T^2 + 3\delta\mu^6\right),
    \end{split} 
\end{equation}
where
\begin{equation}
    \theta_p = \begin{cases}
        0 & k_{Fn} > 3k_{Fp} + k_{Fe} \\
        \dfrac{(3k_{Fp} + k_{Fe} - k_{Fn})^2}{k_{Fn}k_{Fe}} & 3k_{Fp} + k_{Fe} > k_{Fn} > 3k_{Fp} - k_{Fe} \\
        \dfrac{4(3k_{Fp} - k_{Fn})}{k_{Fn}} & 3k_{Fp} - k_{Fe} > k_{Fn} > k_{Fp} + k_{Fe} \\
        2 + \dfrac{2(3k_{Fp} - k_{Fn})}{k_{Fe}} - \dfrac{3(k_{Fp} - k_{Fe})^2}{k_{Fn}k_{Fe}} & k_{Fn} < k_{Fp} + k_{Fe}.
    \end{cases} 
\end{equation}

\section{non-leptonic strange rates}\label{sec:non-leptonic}
For the non-leptonic rates \eqref{reac:non-leptonic}, the rates are computed according to Appendix A of \citet{Ghosh:2026ldg}, with the net rate 
\begin{equation}
    \Gamma_\Lambda = \Gamma_{nn\leftrightarrow n\Lambda} + \Gamma_{np\leftrightarrow \Lambda p} + \Gamma_{n\Lambda\leftrightarrow\Lambda\Lambda},
\end{equation}
with each partial net rate for a reaction $1 + 2 \longleftrightarrow 3 + 4$ given by
\begin{equation}
    \Gamma_{12\leftrightarrow34} = \frac{1}{6}\frac{I_{1234}}{(2\pi)^82^4S}\mu_S\left(4\pi^2T^2 + \mu_S^2\right),
\end{equation}
where $I_{1234}$ is an angular integral over the matrix element and the momentum-conserving delta function, and $S$ is a symmetry factor (see \citet{Ghosh:2026ldg} for details).

\section{Validation of the reaction network}
\label{subsec:validation}
\begin{figure}
    \centering
    \includegraphics[width=0.48\linewidth]{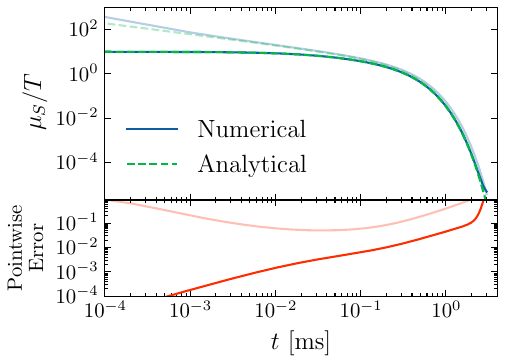}
        \includegraphics[width=0.48\linewidth]{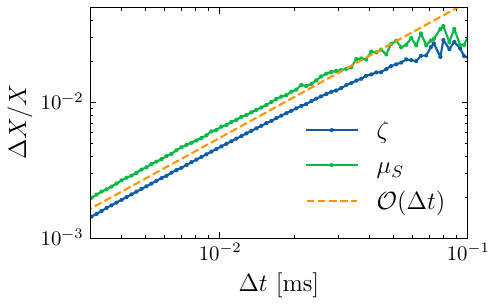}
    \caption{{\it (Left)} Comparison of numerical and analytical solutions for a single equilibrating reaction and the pointwise relative error between the two solutions. {The faint lines show the case of an initial perturbation $\mu_S/T = 10^3$, while the opaque lines show $\mu_S/T = 10^2$.} {\it (Right)} Relative error of extracted bulk viscosity and strangeness chemical potential at a fiducial time as a function of timestep. The orange dashed line corresponds to the expected linear convergence.}
    \label{fig:analyticalcomparison}
\end{figure}

To validate the reaction network evolution, we show in the left panel of Figure \ref{fig:analyticalcomparison} the evolutions of two suprathermal perturbations, $\delta\mu_S / T = 10^2$ and $10^3$, at an initial temperature $T = \SI{10}{\kilo\eV}$, including only the reaction $n + p \longleftrightarrow \Lambda + p$, and use the timestep $\Delta t = \SI{e-7}{\second}$. By writing (see Appendix \ref{sec:non-leptonic})
\begin{equation}
\begin{aligned}
    \frac{d\mu_S}{dt} &= \left.\frac{\partial\mu_S}{\partial Y_S}\right|_{n_B, Y_Q, \varepsilon}n_B \frac{d\delta Y_S}{dt} = -\left.\frac{\partial\mu_S}{\partial Y_S}\right|_{n_B, Y_Q, \varepsilon}\Gamma_{np\leftrightarrow\Lambda p} \\ 
    &= -A\left(4\pi^2\frac{\mu_S}{T} + \left(\frac{ \mu_S}{T}\right)^3\right)
\end{aligned}
\end{equation}
and assuming $A$ and $T$ remain constant over the evolution, we obtain the analytical trajectory in Figure \ref{fig:analyticalcomparison}, which qualitatively agrees well with our numerical results. We also show the pointwise relative error $|( \mu_S/T)_\mathrm{num}- ( \mu_S/T)_\mathrm{ana}|/( \mu_S / T)_{\mathrm{ana}}$ to gauge the difference between the two solutions. For the initial perturbation of $\mu_S/T=10^2$, the error grows over time, and ultimately decreases as $|\mu_S|$ approaches the numerical tolerance. 
For the case of $\mu_S/T = 10^3$, the error is initially large, due to the initial large reaction rates, but decreases during the suprathermal ($\mu_S/T >1$) decay. In both cases, we note that the observed slower decay of the numerical solution is expected using the backward Euler method.

Since the bulk viscosity extracted using Equation \eqref{eq:NSlimit} is of importance to our analysis, we examine its dependence on the employed timestep. In the right panel of Figure \ref{fig:analyticalcomparison}, we show the error convergence in $\mu_S(t_0)$, at a fiducial time $t_0$, and bulk viscosity with decreasing timestep for a $\SI{1}{\kilo\hertz}$ sinusoidal density oscillation at $n_B = 3.5\,n_\mathrm{sat}$ and initial temperature $T = \SI{10}{\kilo\eV}$. We use a reference solution computed with $\Delta t = \SI{e-7}{\second}$, and find that both quantities converge to first order, as expected from the first-order implicit Euler method. 

Lastly, we ensure the conservation of baryon density and total charge density during a number of equilibration tests (at constant baryon density). We find satisfactory conservation of these quantities, with maximal deviations $\Delta n_B / n_B = 9.2\cdot 10^{-10}$, $\Delta n_Q / n_B = 5.0\cdot 10^{-14}$, where $n_Q$ is the total charge density, including both electrons and baryons.

\bibliography{references,Elias_inspire}{}
\bibliographystyle{aasjournalv7.1}

\end{document}